\documentclass[
 prb,
 onecolumn,
 amsmath,amssymb,
 aps,
,citeautoscript,nofootinbib]{revtex4-2}
\usepackage[inkscapelatex=false]{svg}
\usepackage{booktabs} 

\usepackage{graphicx}
\usepackage{setspace}
\usepackage{dcolumn}
\usepackage{bm}

\usepackage{xcolor}
\usepackage{amsmath} 

\usepackage{titlesec}

\usepackage{siunitx}
\begin{document}

\preprint{APS/123-QED}
\title{High Harmonic Spectroscopy from Lower-Order to  Higher-Order Topological Insulators}

\author{Bryan Lorenzo$^{1}$}
\author{Carlos Batista$^{1}$}
\author{Milad Jangjan$^{2}$}
\author{Dasol Kim$^{3}$}
\author{Jean Menotti$^{1}$}
\author{Feng Liu$^{4}$}
\author{Wenlong Gao$^{4}$}
\author{Shambhu Ghimire$^{5}$}
\author{Camilo Granados$^{4}$}
\author{Alexis Chac\'on$^{1,6,7,8}$}
\email{alexis.chacon-s@up.ac.pa }

\affiliation{$^{1}$Departamento de F\'isica, \'Area de F\'isica, Universidad de Panam\'a,
Ciudad Universitaria 3366 Octavio Mendez Pereira, Panama}
\affiliation{$^{2}$Institute of Physics, University of Rostock, 18051 Rostock, Germany}
\affiliation{$^{3}$Department of Materials Science and Engineering, Pohang University of Science and Technology, Pohang 37673, Korea}
\affiliation{$^{4}$Eastern Institute of Technology, 315200, Ningbo, China}
\affiliation{$^{5}$ Stanford PULSE Institute, SLAC National Accelerator Laboratory,
Menlo Park, California 94025, United States}
\affiliation{$^{6}$Sistema Nacional de Investigación de Panamá,  Building 205,
Ciudad del Saber ,Clayton Panamá, Panama}
\affiliation{$^{7}$Centro de Investigaci\'on con T\'ecnicas Nucleares,
Universidad de Panam\'a, Panama}
\affiliation{$^{8}$Parque Cient\'ifico y Tecnol\'ogico, Universidad Aut\'onoma de Chiriqu\'i, 
Ciudad Universitaria, David, Panama}

\date{\today}

\begin{abstract}
{\color{black}
Over the past decades, high-harmonic spectroscopy (HHS) has emerged as a powerful tool for all-optical probing of topological properties of solids.~There are outstanding questions regarding universal nature of the spectral features of harmonics in their relationship to the non-trivial topological properties.~Here, we present a systematic theoretical study of HHS in topological materials, including lower-order and higher-order topological insulators (LOTIs and HOTIs), focusing on observables such as {\it helicity, circular dichroism, ellipticity dependence}, and {\it channel-resolved intensity yields}.~Using the Haldane, Kane–Mele, and breathing Kagome lattice models, we theoretically extend all-optical approaches from the LOTI to the HOTI regime by explicitly incorporating contributions from {\it bulk, edge}, and {\it corner states}.~Depending on the crystalline system, our calculations suggest that these observables can  encode topological information through distinct modifications of the HHG spectra in topological phases.~In particular, we identify significant {\it enhancements} of the harmonic intensity yields, reaching up to two orders of magnitude relative to trivial phases, together with distinct spectral signatures associated with {\it edge} and {\it corner} contributions revealed through channel-resolved intensity yields.~These results show that channel-resolved HHS provides a promising route for probing topological states in both LOTIs and HOTIs.}
\end{abstract}

\maketitle
\doublespacing 

\textit{Introduction}

{\color{black}Topological physics is a unique branch of condensed matter physics devoted to the study of topological phases and transitions, which are governed by symmetries and topological invariants~\cite{HasanRev.Mod.Phys2010,Xiao-LiangQiRevModPhys2011,ThoulessPRL1982}. These phases are of great importance not only for fundamental science but also for the development of advanced technological devices, such as topological transistors~\cite{FleetNatPhy2015,GilbertCommPhy2021}. Materials in this class are distinguished by the presence of conducting topological {\it states at their edges or surfaces}, while remaining {\it insulating in the bulk}~\cite{Kane1PRL2005,Kane2PRL2005,MarkusScience2007,bernevigScience2006,FuPRL2007,XiaNaturePhysics2009}.
This remarkable duality is protected by symmetries and characterized by topological invariants, i.e., the Chern number or $\mathbb{Z}_2$ invariants. Topological materials can be broadly classified into lower-order topological insulators (LOTIs)~\cite{Kane1PRL2005,Kane2PRL2005,Xiao-LiangQiRevModPhys2011,ThoulessPRL1982,HasanRev.Mod.Phys2010} and higher-order topological insulators (HOTIs)~\cite{BenalcazarPRB2017,BenalcazarScience2017,SchindlerScienceAdv2018,SongPRL2017}.~LOTIs are characterized by a gapped bulk and gapless boundary states~\cite{HasanRev.Mod.Phys2010,Xiao-LiangQiRevModPhys2011}.~More generally, for an $n$-dimensional LOTI, the {\it bulk} is an insulating $n$D phase, while the ($n-1$)D boundary hosts topologically protected conducting states (here, $n$ denotes the dimensionality of the system). These boundary modes are protected by topological invariants, such as the Chern number and the $\mathbb{Z}_2$ invariants, which arise as a direct consequence of the system symmetries.}

\begin{figure}
    \centering
    \includegraphics[width=0.95\linewidth]{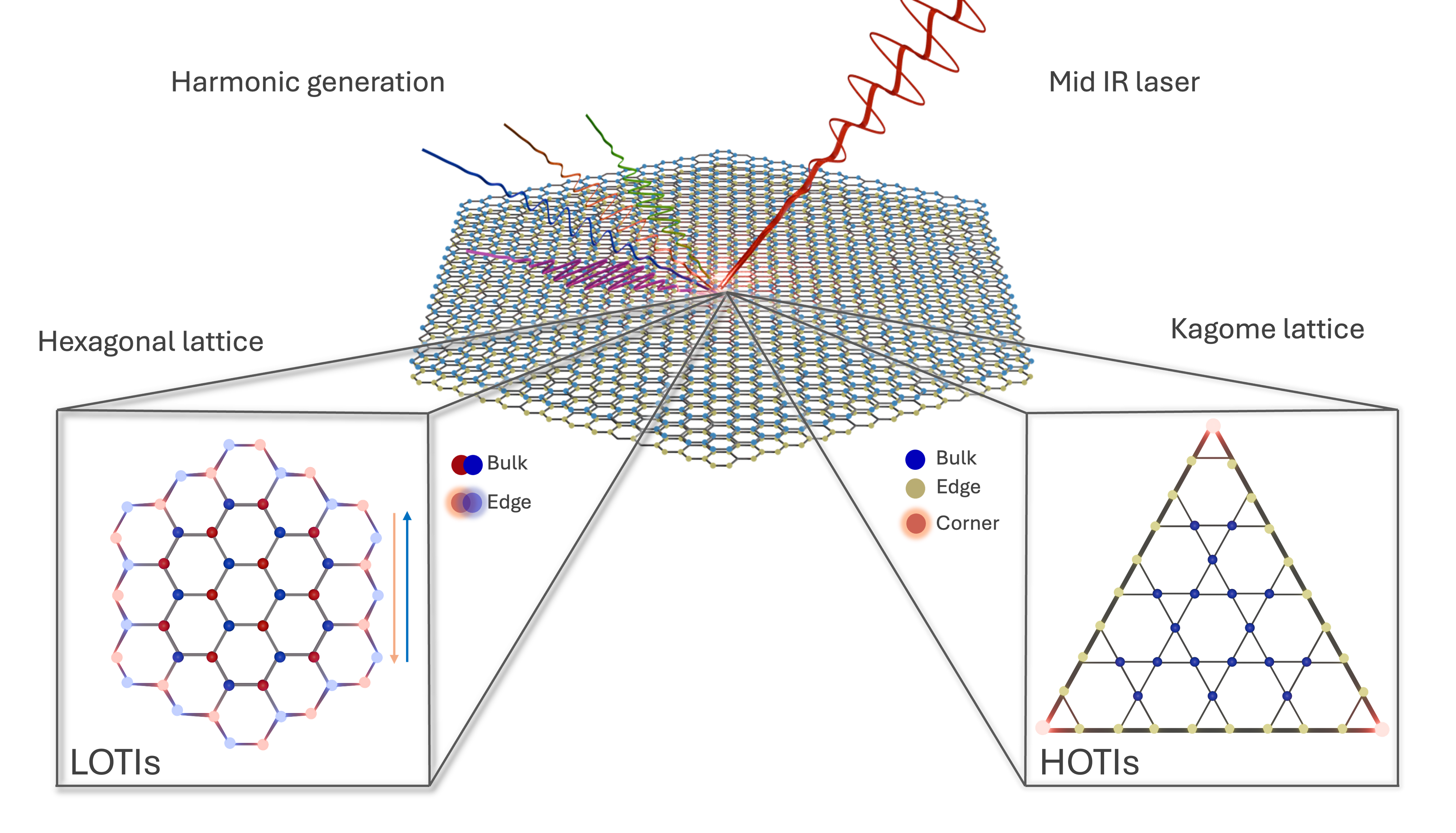}
    \caption{High-harmonic generation in two-dimensional topological materials:~A strong and ultrashort mid infrared (MIR) driving laser~(red pulse) is focused on a topological honeycomb lattice.~As a result of the laser–matter interaction, high-energy photons (illustrated by oscillating green, orange, blue, and violet lines) are emitted from the systems considered here:~lower-order topological insulators {\color{black}(LOTIs)}, including 2D~Chern insulators described by the Haldane model, which breaks both inversion and time-reversal symmetries, and 2D topological insulators (TIs) modeled by the Kane–Mele (the two atoms of the unit cell are in blue and red dots), which preserves time-reversal symmetry~(see left {\color{black}hexagonal inset zoom}). In the case of the 2D TI, the conducting topological edge states are represented (shadow red-blue dots) orange and blue arrows along the edges of the hexagonal lattice,~and the light red and blue circles represent those conducting states.~The~right-hand crystalline structure is the higher-order topological insulators {\color{black}(HOTIs)} based on a breathing Kagome lattice ({\color{black}red shadow circles}) hosting topological corner states ({\color{black}right triangular inset zoom}), where the conducting states are visualized as red illuminated points, the edges as yellow and the bulk as blue circles.}
    \label{fig:fig0}
\end{figure}

\noindent In particular, topological insulators (TIs) exhibit {\it gapless edge} or surface states coexisting with a {\it gapped bulk}. Here, we focus on two-dimensional (2D) TIs unless otherwise stated, where conducting channels appear at the one-dimensional (1D) edges, while the 2D bulk remains insulating.~Representative examples of LOTIs include the Haldane model~\cite{HaldanePRL1988}, describing Chern insulators (CIs); the Kane--Mele model~\cite{Kane1PRL2005,Kane2PRL2005,bernevigScience2006,MarkusScience2007}, which realizes 2D TIs; and the Zhang model for Bi$_2$Se$_3$~\cite{ZhangNatPhy2009}, a paradigmatic 3D TI. As a concrete example, a 2D TI such as 1T'-WS$_2$ hosts {\it one-dimensional} (1D) {\it conducting edge states} while remaining {\it insulating} in the 2D {\it bulk}, as illustrated in Fig.~\ref{fig:fig0} (hexagonal lattice).
On the other hand, HOTIs are generally defined as topological $n$-dimensional materials whose {\it insulating states} reside in $n$ dimensions (identified as the {\it bulk states}), while their {\it topological states} appear in $(n-2)$D (identified as the {\it corner states}) ({\color{black}see {\it corner states} in Fig.~\ref{fig:fig0} for the Kagome lattice}). The Kagome lattice can be considered either a LOTI or a HOTI (2D material)~\cite{GuoPRB2009,BolensPRB2019,EzawaPRL2018,KempkesNatureMat2019,LiuPRA2010}, depending on the parameters of the Hamiltonian model (see Fig.~\ref{fig:fig0} for the Kagome lattice). It exhibits topological corner states, where the {\it corners} correspond to zero-dimensional (point-like $0$D) states~\cite{EzawaPRL2018,KempkesNatureMat2019,HerreraPRB2022}, while the ($n-1$)D {\it edges} and the 2D {\it bulk} remain insulating~\cite{OhgushiPRB2000,BolensPRB2019,LiuPRA2010,GuoPRB2009}.
Some of the materials described above and experimentally realized, such as Chern insulators~\cite{ZhaoNaturePhysics2024}, topological insulators (TIs)~\cite{MarkusScience2007}, and higher-order topological insulators~\cite{KempkesNatureMat2019}, belong to a broader class of topological phases known as symmetry-protected topological (SPT) materials~\cite{Xiao-LiangQiRevModPhys2011,BenalcazarScience2017}. Certain topological phases rely on the presence of protecting symmetries (e.g., time-reversal, inversion, or mirror symmetry). When these symmetries are broken, the associated topological features may disappear. For example, in 3D topological insulators such as Bi$_2$Se$_3$, impurities (e.g., Indium doping) can suppress the topological surface states by modifying the electronic structure and effectively breaking the protecting symmetry~\cite{HasanRev.Mod.Phys2010,BrahlekPRL2012,SalehiNanoLetters2016}. In contrast, in Chern insulators the topological phase emerges from the explicit breaking of inversion and time-reversal symmetries rather than from a protecting symmetry.

\noindent In this work, we {\color{black}focus on} the study of lower-order and higher-order topological insulators through the nonlinear optical response of these materials.~{\color{black}Our targets are described using 2D topological models such as the paradigmatic Haldane model (2D Chern insulator~\cite{JotzuNature2014,HaldanePRL1988}), the Kane–Mele model (2D TI)~\cite{Kane1PRL2005,Kane2PRL2005} for LOTIs, and the Kagome-lattice structure (2D triangular lattice) for HOTIs~\cite{EzawaPRL2018} (see Fig.~\ref{fig:fig0}).}~The Kagome lattice, due to its geometric frustration, spatial symmetry, and band topology, serves as an ideal platform for the realization of HOTIs~\cite{EzawaPRL2018}.~{\color{black}For example, HOTIs in Kagome lattices exhibit localized {\it corner states}, or zero-dimensional topological states (see Fig.~\ref{fig:fig0} for the Kagome lattice), protected by crystalline symmetries (e.g.,~$\rm C_3$, inversion, etc.)~\cite{EzawaPRL2018,HerreraPRB2022,KempkesNatureMat2019}.}
{\color{black}These states are topologically robust; they arise from quantized multipole moments and crystalline symmetry protection~\cite{BenalcazarScience2017}. Models such as the Ezawa Kagome model demonstrate that these systems host quantized corner charges, but without anyonic excitations or degenerate ground states~\cite{SchindlerNaturePhysics2018}. Hence, Kagome LOTIs and HOTIs belong to the SPT class—specifically as higher-order topological insulators and not to the class of topologically ordered phases~\cite{WENIntlJourModernPhysicsB1990,HasanRev.Mod.Phys2010}.}


\noindent Despite the great success of Angle-Resolved Photoemission Spectroscopy (ARPES) in measuring the energy band structures of TIs~\cite{TangNaturePhy2017,ChenScience2009,HsiehNature2008}, identifying or measuring topological invariants remains a challenge in condensed matter physics~\cite{HasanRev.Mod.Phys2010}. Thus, any alternative technique that can elucidate and address the challenge of characterizing topological materials is highly valuable. {\color{black}In this context, high-harmonic spectroscopy (HHS), based on high-harmonic generation (HHG), is emerging {\color{black} as an alternative nonlinear optical technique} to ARPES measurements for exploring topological materials, their phases, and their transitions~\cite{BauerPRL2018, BauerPRB2019, ChaconPRB2020, BaykushevaNanoLetters2021, BaykushevaPRA2021, HeideNaturePhotonics2022, MitraNature2024}.}~High-harmonic generation is a nonlinear optical process in which the photon energy of a laser focused on a medium (gas, solid, or liquid) is upconverted into high-frequency photons with energies that are integer multiples of the fundamental laser frequency~\cite{LewensteinPRA1994}, as it is shown in Fig.~\ref{fig:fig0} for the optical response from solids~\cite{GangarajPRR2020}.~This process occurs on a sub-femtosecond timescale in the atoms of a gas or solid and captures rich information about the electronic structure of the target via dipole transition matrix elements or induced currents, energy bands of the crystal, etc.~\cite{VampaPRL2015,BaykushevaPRA2021,BaykushevaNanoLetters2021}.  Recently, HHG has been theoretically investigated as a probe of topological phases and transitions in systems such as the Su–Schrieffer–Heeger (SSH) model~\cite{BauerPRL2018}, the Haldane model (Chern insulator)~\cite{JotzuNature2014}, the Weyl semimetals~\cite{Gopal1PRB2023,Gopal2PRB2023, Gopal3PRB2023,LvNatureComuni2021,LiuPRB2025,AvetissianPRA2022,ZhangarXiv2024} and three-dimensional (3D) topological insulators like Bi$_2$Se$_3$~\cite{BaykushevaNanoLetters2021,BaykushevaPRA2021}, to mention some. by analyzing the  helicity-resolved asymmetry or degree of circular polarization, the circular dichroism, and {\color{black}the anomalous ellipticity dependence} of the emitted harmonics{~\cite{ChaconPRB2020,HeideNaturePhotonics2022, KimMDPI2022,AtsushiPRB2024}. 

\noindent Notwithstanding these important experimental demonstrations, theoretical calculations of HHS by Neufeld {\it et al.} suggest that the HHG mechanism does not always encode topological information as a universal probe~\cite{NeufeldPRX2023}. Their study focused on BiH and Na$_3$Bi–like materials using {\it ab-initio} time-dependent density functional theory (TDDFT).~{\color{black}However, this discrepancy may not be universal across all topological materials. In particular, we expect that the full high-harmonic spectrum originates not only from {\it topological bulk states}, but also from contributions of {\it topological edge states {\rm (or} surface states{\rm)}}, as well as {\it topological corner states}~\cite{BauerPRL2018,HeideNaturePhotonics2022,BaykushevaNanoLetters2021,SchmidNature2021}.}

These states can play a crucial role in shaping topological phases and transitions in HHS, even in cases where such emission channels were not considered in Ref.~\cite{NeufeldPRX2023}, as demonstrated experimentally by Baykusheva~{\it et al.}~\cite{BaykushevaNanoLetters2021} and Heide~{\it et al.}~\cite{HeideNaturePhotonics2022}, respectively.~{\color{black}For example, by analogy with quantum Hall phenomena,} in the integer quantum Hall effect (IQHE)~\cite{KlausRMP1986} and in the fractional quantum Hall effect (FQHE)~\cite{StormerRMP1999}, the {\it charge current} is the key observable that encodes the underlying topology. In contrast, in quantum spin Hall systems (QSHE), the {\it spin current} is the relevant measurable quantity that captures the topological properties of the system~\cite{KlitzingPRL1980,StormerRMP1999,Kane1PRL2005}.~A similar situation is expected for HHS, where the emitted harmonics may serve {\color{black}as sensitive probes of topology in condensed matter systems.}
Baykusheva~{\it et al.}~\cite{BaykushevaNanoLetters2021} and Heide~{\it et al.}~\cite{HeideNaturePhotonics2022} presented theoretical calculations and measurements indicating that the high harmonics originating from the {\color{black}topological {\it surface states} of a 3D TI Bi$_2$Se$_3$ exhibit an {\it atypical, or anomalous, ellipticity dependence} (AED)}.~{\color{black}Other experimental measurements have indicated that HHS exhibits two-dimensional topological surface-wave radiation (BiSbTeSe$_3$), THz emission associated with fractional harmonics ($\rm Bi_2Te_3$), and clear enhancements in topologically nontrivial phases relative to trivial ones
\cite{BaiNatPhys2021,SchmidNature2021,TielrooijLSA2022}}.
This discrepancy has sparked an ongoing debate in the attosecond-physics and condensed-matter physics communities~\cite{NeufeldPRX2023,BaykushevaNanoLetters2021,BaykushevaPRA2021,ChaconPRB2020,JimenezGalanNaturePhotonics2020,MitraNature2024}.

{\color{black}Here, we aim to explore high-order harmonics generated from LOTIs to HOTIs using high-harmonic spectroscopy (HHS), to address this debate. Three distinct topological models, characterized by different topological invariants ($C_m$, $Z_2$, and $P_3$, see Appendix~\ref{AppendixA}) and symmetry protections, are studied to elucidate whether and how the HHG process encodes 
topological information in LOTIs and HOTIs~\cite{HaldanePRL1988,ThoulessPRL1982,Kane2PRL2005,BenalcazarPRB2017}.~We analyze channel-resolved intensity yields together with helicity, circular dichroism, and AED for harmonic emissions originating from bulk, edge, and topological corner states in the corresponding LOTI and HOTI systems.~In the case of LOTIs, these studies are addressed not only through high-harmonic emissions from the topological {\it bulk states}, but also from the topological {\it edge states} in 2D TIs and 2D CIs.~In the case of HOTIs, we compute the full nonlinear all-optical response from the topological {\it bulk}, {\it edge}, and {\it corner states}.~Our theoretical results suggest that HHS is a useful technique to probe topological materials, i.e., LOTIs and HOTIs.~For the topological Chern insulators, for example, the  helicity-resolved asymmetry or degree of circular polarization~\cite{JimenezGalanNaturePhotonics2020} and the circular dichroism of the emitted high-order harmonics~\cite{ChaconPRB2020} (specifically in the Haldane model; see Appendix~\ref{AppendixA.1} for details) indicate {\it qualitative} differences between topologically trivial and nontrivial phases~\cite{BauerPRL2018,ChaconPRB2020,BaykushevaPRA2021}. In the case of TIs described by the 2D Kane-Mele model, the HHG intensity yield as a function of the driving-laser ellipticity {\it exhibits an atypical tendency}, in direct agreement with the experimental observations reported in Ref.~\cite{BaykushevaNanoLetters2021} and theoretical calculations~\cite{BaykushevaPRA2021}. {\color{black}This quantity is referred to in this work as {\it anomalous ellipticity dependence} (AED).~In particular, we further extend our HHG calculations to investigate the emitted radiation from {\it topological edge states} in our LOTI models.~A clear intensity enhancement of the emitted harmonics is observed in the simulated HHG spectra from the topological {\it edge states}, compared to both the topological {\it non-trivial bulk states} and the {\it trivial bulk states}, for the Haldane and Kane--Mele models in analogy with the investigation carried out by Baykusheva {\it et al.} in Ref.~\cite{BaykushevaPRA2021}.}~This suggests a general trend in LOTI materials that cannot be explained without considering all possible emission channels, namely bulk and edge contributions. For the HOTI model considered in this work (the {\it breathing Kagome lattice}), we observe signatures of topology in the high-harmonic yield from the {\it topological corner states}, in comparison with the emissions from the {\it bulk states} and the {\it edge states}.
The electronic dynamics for the topological bulk states, edge states and corner states are addressed by solving both: (i) the time-dependent density matrix~\cite{KimMDPI2022},~(see Appendix.~\ref{AppendixB}) and (ii) Bauer’s theory~\cite{BauerPRB2019,BauerPRL2018}, which is partially based on the SSH model (see Appendix~\ref{AppendixB.2}).~While the calculated HHG from bulk and edge states is robust, the emitted harmonics from the {\it corner states} are extremely sensitive to the parameters of the SSH model.~We present HHG calculations in which a clear difference in the harmonic intensity yields between trivial and nontrivial phases is observed for the corner states; however, this feature does not persist across all model parameter regimes.~This indicates that HHS can  
encode topological information through harmonic intensity yields, albeit in a material-dependent manner.~Finally, we emphasize the significant enhancement, exceeding two orders of magnitude, in the high harmonics emitted from the topological Kagome lattice in the semimetallic phase.}}~In this context, four HHG-based observables are particularly useful for probing topological phases and transitions~\cite{BaykushevaNanoLetters2021,HeideNaturePhotonics2022}, namely:
(i)~Helicity, the asymmetry between harmonics emitted with right- and left-handed circular polarization when driven by a {\it linearly polarized} infrared laser or helicity-resolved asymmetry;
(ii)~Circular dichroism, the normalized intensity difference between harmonic orders generated by right- and left-handed {\it circularly polarized} driving lasers;
(iii)~Ellipticity dependence, the variation of harmonic yield as a function of the driving laser ellipticity and {\color{black} (iv) the relative HHG intensity yield of harmonics emitted from {\it topological edge states} and {\it corner states}, in comparison to {\it bulk states} (trivial or non-trivial topological phases).
In our theoretical analysis, we employ all four observables, unless otherwise stated.}

\textit{Lower-order topological insulators}\\
{\color{black}As a benchmark for validating our models, in particular the emission from topological bulk and edge states, and probing topological signatures in high-harmonic spectroscopy,} we first analyze the nonlinear all-optical response of lower-order topological insulators (LOTIs), specifically the two-dimensional Chern insulator and topological insulator, modeled by the Haldane and Kane--Mele Hamiltonians, respectively (see Appendix~\ref{AppendixA} for details). {\color{black}The choice of parameters for the trivial and non-trivial phases was guided by realistic energy scales motivated by Kagome materials such as Nb$_3$Cl$_8$, Nb$_3$Br$_8$, and Nb$_3$TeCl$_7$~\cite{KMTBM, SunNanoLett2022, ZhangAdM2023}, whose band gap is approximately~$\sim 1.0$~eV, allowing for a consistent theoretical comparison with the Kagome-based models.}
{\color{black}Figures~\ref{fig1}a) and \ref{fig1}e)~(\ref{fig1}c) and \ref{fig1}g)) present the energy dispersions for the trivial and non-trivial topological phases of the Haldane model (Kane--Mele model) for 2D bulk states.~The corresponding HHG spectra for the trivial and nontrivial topological phases of the Haldane model (Kane--Mele model) are shown in Figs.~\ref{fig1}b) and \ref{fig1}f) (\ref{fig1}d) and \ref{fig1}h)), respectively.}


\noindent When the laser pulses: (i)~linearly polarized (LP), (ii)~right-handed circularly polarized (RCP), and (iii)~left-handed circularly polarized (LCP) drive the topological Haldane model in its trivial phase, no substantial difference is observed in the high harmonics generated from the Chern insulator (see Fig.~\ref{fig1}b)).~{\color{black} Note, however, that the HHG spectra produced under LP driving from a non-trivial topological phase exhibit a significantly enhanced compared to the trivial phase, typically by two to three orders of magnitude. These harmonics are calculated for the topological trivial (non-trivial) bulk states. {\color{black}Below, we extend our model from topological {\it bulk} harmonic emissions to {\it topological edge} and {\it topological corner states}.}~These harmonics fully follow the selection rules corresponding to the ${\rm C}_3$ symmetry and broken inversion symmetry of the Haldane model. This means that we can observe the harmonic selection rules for LP driving, {i.e.}, $n' : 1, 2, 3, \dots$, where both even and odd harmonic orders (HOs) are allowed. The harmonics produced by RCP and LCP lasers also exhibit the three-fold selection rule given by $3n' \pm 1$. We further observe that the harmonic intensities for the trivial phase are almost the same for these driving conditions. In contrast, in the topological phase, an asymmetry emerges in the harmonics across the HHG spectral plateau (see Fig.~\ref{fig1}f)), particularly for the co-rotating HOs.

\begin{figure}[h!]
    \centering 
    \includegraphics[width=1\textwidth]{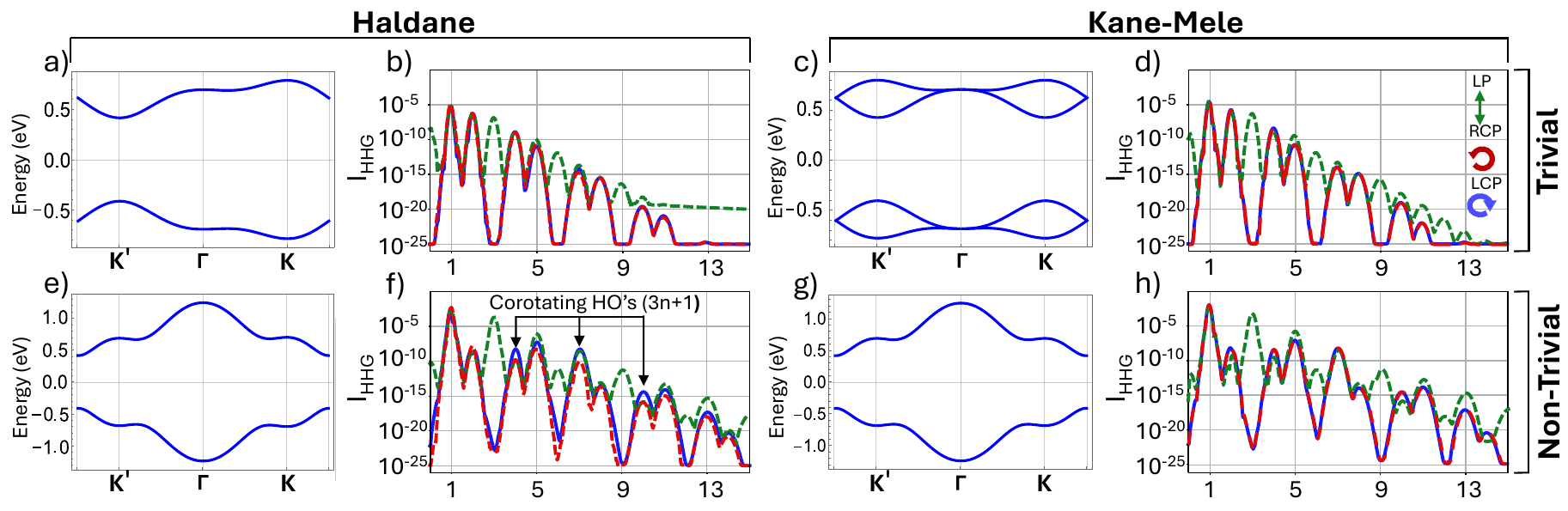}
    \caption{ {\color{black}Band dispersions and high-harmonic generation (HHG) for lower-order topological insulators:} Chern insulators and topological insulators.~For the trivial phase~(Chern number, $C = 0$), panels a) and b) show the band structure across the K'--$\Gamma$--K points of the BZ for the Chern insulator (Haldane model) and the corresponding HHG spectra produced by a linearly polarized (LP, green line) laser, right-handed circularly polarized (RCP, red dashed line) laser, and left-handed circularly polarized (LCP, blue line) laser. The parameters used for the trivial phase are: lattice constant \(a_0 \sim 7.0\, \text{\AA}\), nearest-neighbor hopping \(t_1 = 0.0043\)~a.u.~($\sim$ 0.117~{\rm eV}), next-nearest-neighbor hopping \(t_2 = 0.00132\)~a.u.~($\sim \,$0.0359~eV), onsite potential ratio \(M/t_2 = 16.77\) and magnetic flux  \(\phi_0 = \pi/2\). Similarly, for the topological phase ($C = +1$) in the Haldane model, panels e) and f) display the topological band structure and the corresponding HHG spectra. The parameters for the Haldane topological phase are: \(t_1 = 0.0152\)~a.u.~($\sim \,$0.41~eV), \(t_2 = 0.0049\)~a.u.~($\sim \,$0.133~eV), \(M/t_2 = 0.0520\) and we keep the same lattice constant and magnetic flux of the trivial phase. Additional calculations for the band structures and HHG spectra are shown in panels c) (four bands are observed) and d) for the trivial phase, and in panels g) (topologically degenerated four bands are drawn) and h) for the non-trivial (topological) phase of the Kane-Mele model, using the same parameters as above for the respective phases, with the Rashba coupling set to zero \(t_R = 0.0\) (see Appendix~\ref{AppendixA.1a} for details of the model).~The laser parameters are field strength \(E_0 = 0.0007\)~a.u.,~intensity \(I_0 \sim
 1.7 \times 10^{10}\)~W/cm\(^2\), photon energy \(\hbar \omega = 0.013\)~a.u.~($\sim$
0.35~eV or 3.54~$\mu$m), and a Gaussian laser field envelope with a pulse duration of 7~opt.~cycles at full width half maximum (FWHM). Atomic units are used throughout this work, unless otherwise stated.~{\color{black}All parameters were chosen such that the band structures of the LOTI models exhibit a band gap of approximately $\sim0.8$~eV, consistent with the reported bandgap of the Kagome-lattice compound Nb$_3$TeCl$_7$.}}
\label{fig1}
\end{figure}



\begin{figure}
    \centering
    \includegraphics[width=0.85\linewidth]{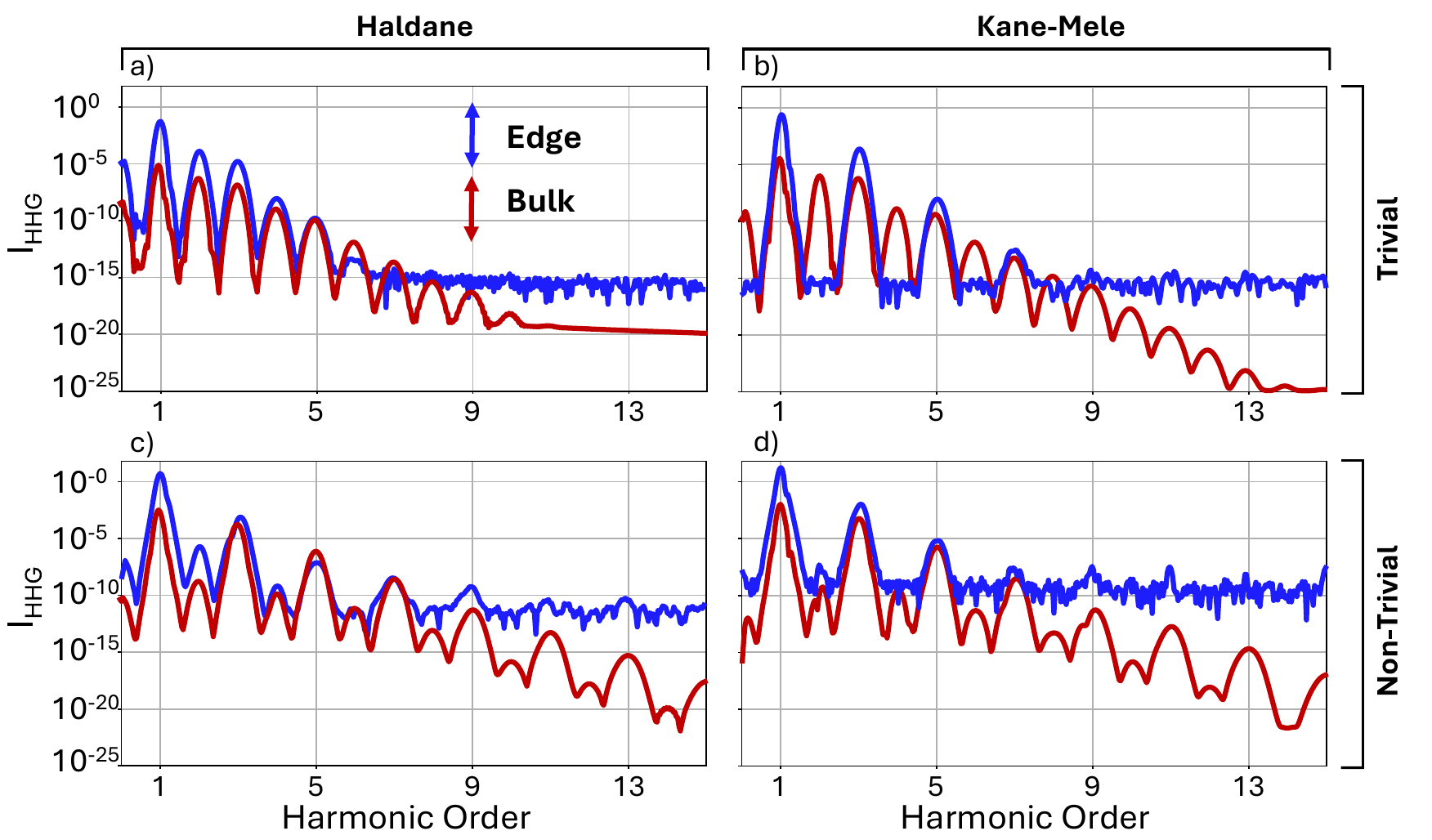}
    \caption{{\color{black}  Comparison of the high harmonic generation (HHG) spectra produced from the {\it topological bulk states} and the {\it topological edge states} in LOTIs:
Panels~a)~and~c) show the HHG spectra for the 2D~Haldane model subjected to a linearly polarized driving laser for the {\it trivial} and the {\it non-trivial} topological phases, respectively.~We compare the intensity of the {\it bulk emission} (red line) with the {\it edge emission} (blue line).~The emission from topological {\it edge states} shows an interesting enhancement between {\it one and two} orders of magnitude over the topological {\it bulk states}.~Panels b) and d) exhibit the same analysis as in panels a) and c), but now for the 2D topological insulator described by the Kane--Mele model.~All simulations (bulk and edge) use the same linearly polarized laser and topological material parameters as in Fig.~\ref{fig1}.~Edge states are computed for an one-dimensional zig-zag hexagonal strip~(See Appendix~\ref{AppendixA.1b}).~Both the Haldane and Kane--Mele models include next-nearest-neighbor terms, whereas only the Kane--Mele model incorporates spin (up and down).~We note that the high-order harmonic intensities emitted from the topological {\it edge states} are {\it enhanced} in comparison to the topological {\it bulk states} for our LOTI models. }}
    \label{fig:fig2}
\end{figure}


\begin{figure}
    \centering
    \includegraphics[width=0.9\linewidth]{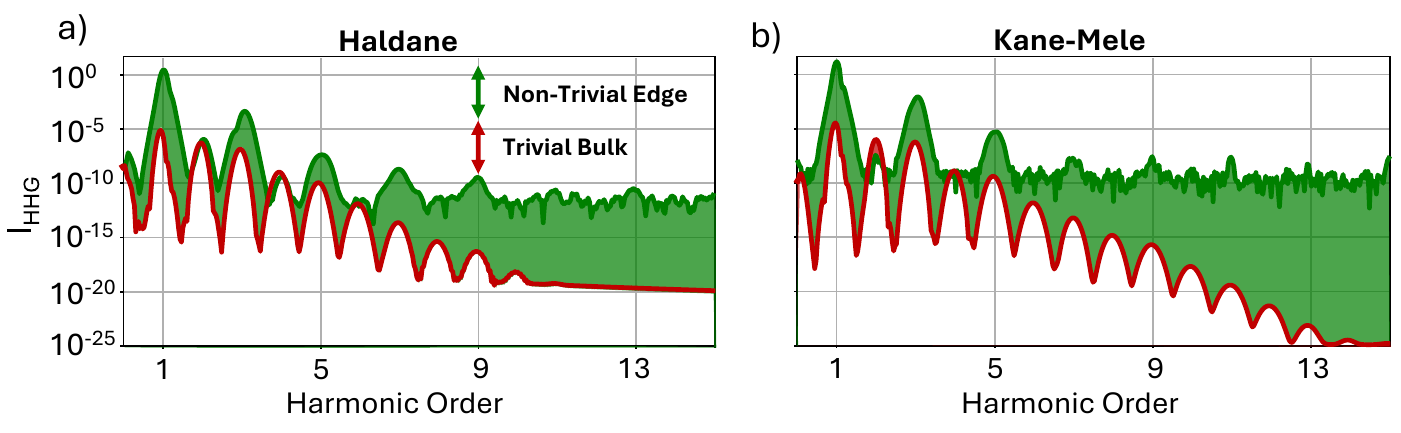}
    \caption{{\color{black} Comparison of the high-harmonic generation (HHG) spectra for the Haldane and Kane--Mele models: Panel~a) shows the HHG spectra for the 2D Haldane model driven by a linearly polarized laser, comparing emission from the {\it trivial bulk states} (red line) and the {\it non-trivial edge states} (green line).~Panel~b) presents the same analysis for the Kane--Mele model, also contrasting the {\it trivial bulk emission} and the {\it topological edge emission}.~The laser and material parameters are the same as in Fig.~\ref{fig:fig2}.~The harmonic intensity yield for both materials (Haldane and Kane–Mele models) exhibits an enhancement of about two to four orders of magnitude for topological edge emission compared to the trivial case.} }
    \label{fig:fig3}
\end{figure}

\noindent {\color{black}To this point, our calculations presented in this work correspond to the topological bulk states.~\noindent {\color{black}However, a natural question arises: can the inclusion of {\it topological edge (or corner) states} in the model provide a more ``direct" signature of the topology material in the HHG response? To address this question, we extend our models to explicitly include topological {\it edge states} in the nonlinear optical emission. Since the system remains periodic along one spatial direction, we compute the current along that direction, in close analogy with the bulk calculation (see Appendix~\ref{AppendixB.1}).}~Thus, our theoretical framework incorporates the topological {\it conducting edge states} in the calculations of HHG for both the Haldane model and the Kane--Mele model, i.e.,~our LOTI models. We use zigzag nanoribbon models for a hexagonal lattice~\cite{TraversonpjQuatumMat2024}, preserving periodicity along the $x$-direction while restricting the material along the $y$-direction.~This allows us to study the emission from the edge of the material where topological states may exist (see Appendix~\ref{AppendixA.1b}).
We then compare the intensity of the harmonic emissions from the topological {\it bulk states} and the topological {\it edge states}.~The results are shown in Fig.~\ref{fig:fig2} and~Fig.~\ref{fig:fig3}.~For all LOTI models, we find a relatively good enhancement (one or two orders of magnitude) in the HHG intensity yield for the topologically {\it non-trivial phases} compared to the topologically {\it trivial phases}. The enhancement is observed for the emission of the topological edge states (see all curve-lines depicted in blue lines of~Figs.~\ref{fig:fig2}) in comparison with the topological bulk states~(see all Figs.~\ref{fig:fig2} depicted in red lines).~This enhancement suggests a {\it promising} indication that topological {\it conducting edge states} can be distinguished in 2D-TIs by means of HHG spectrum analysis or HHS.~A similar conclusion was shown by Baykusheva {\it et al.}~\cite{BaykushevaNanoLetters2021} in the context of the 3D-TI in Bi$_2$Se$_3$ where the enhancement was observed in the 2D topological {\it surface states} in comparison with the 3D-topological {\it bulk states}~\cite{BaykushevaPRA2021}. 
}

\noindent {\color{black}Additionally, we compare the HHG emission originating from the trivial bulk states with the emission generated from topological edge states (as shown in Fig.~\ref{fig:fig3}).~In particular, the Haldane model (Fig.~\ref{fig:fig3}a) and the Kane--Mele model (Fig.~\ref{fig:fig3}b) are analyzed.~The results show that the emission intensity associated with {\i the topological edge states} significantly exceeds the bulk emission in the trivial phase.~This behavior can be 
attributed to the presence of topologically protected edge states, which exhibit stronger spatial localization and a more efficient coupling with the incident laser field.~Consequently, these results suggest that materials in a topological phase favor more intense nonlinear emission processes than their trivial counterparts.~Our theoretical results suggest that topological signatures are  encoded in high-harmonic generation for lower-order topological insulators.~In particular, this becomes noticeable once we consider the full harmonic emission channels from LOTIs:~the 2D topological insulating {\it bulk states} and the~1D topological conducting~{\it edge states} in both: Haldane model and Kane---Mele model.~{\color{black}Figs.~\ref{fig:fig3}a) and~\ref{fig:fig3}b) further suggest that, for systems sharing the same minimum bulk bandgap, differences in the harmonic intensity yields between trivial and topological phases may provide {\it indirect} signatures of topology, even in systems exhibiting similar gapless edge states. Experimentally, this prediction could be tested by comparing the high-harmonic emission from a trivial reference target (MoS$_2$ \cite{ThomasMaterTodaySust2021}) with that from a 2D topological insulator (Bismuthene, BiC$_2$Br~\cite{ReisScience2017,WuApplSurfSci2019}), selected to possess a similar minimum bandgap.}

\noindent{\color{black}To connect with previously established observables used to probe the topological properties of quantum materials, we now return to bulk-related quantities.~In particular, we consider the  helicity-resolved asymmetry ($h^{(n)}$) and the circular dichroism (CD$^{(n)}$).~First, we define $h^{(n)}$ as the normalized intensity difference ($I_{\rm LP}^{(n,\pm)}$) between the right-handed circularly polarized (RCP, $+$) and left-handed circularly polarized (LCP, $-$) {\it components} of the high harmonics generated by a linearly polarized (LP) driving laser (see Table~\ref{tab1}):}
\begin{eqnarray}
h^{(n)}=\frac{ I_{\rm LP}^{(n,+)} - I_{\rm LP}^{(n,-)} }{ I_{\rm LP}^{(n,+)} + I_{\rm LP}^{(n,-)} }. \label{eqn:eq1}
\end{eqnarray}
\noindent where $n$ is {\color{black}the harmonic order (HO)} with circularly polarized decompositions~(${\pm}$).~Here, we define the normalized helicity as a continuous quantity, rather than in terms of standard chiralities (i.e. $ h^{(n)}=(0,\pm 1)$). Note, the subindex LP means the harmonics are generated by a linearly polarized laser.~{\color{black}We compute the  helicity-resolved asymmetry and present it in Table~\ref{tab1}.~The behavior of $h^{(n)}$ is in good agreement with that reported by Silva {\it et al.}~\cite{SilvaNatPho2019}.}~Second, the circular dichroism (CD) is defined in a similar way than the helicity: we measure the harmonics produced by RCP and LCP lasers separately and compare the difference between corotating HOs with $n = 3n'+1$ as 
\begin{eqnarray}
{\rm CD}^{(n)} = \frac{I_{\rm RCP}^{(n)} - I_{\rm LCP}^{(n)}}{ I_{\rm RCP}^{(n)} + I_{\rm LCP}^{(n)} }.\label{eqn:eq2}
\end{eqnarray}

\noindent For the \(4^\text{th}\), \(7^\text{th}\) and  \(10^\text{th}\) harmonics generated from Haldane model, the  helicity-resolved asymmetry exhibits positive values in the trivial phase and negative values in the non-trivial one (see Table~\ref{tab1}).~However, for the \(10^\text{th}\) corotating harmonic the  helicity-resolved asymmetry is zero.~{\color{black}This sign change in the  helicity-resolved asymmetry (for simplicity we refer helicity-resolved asymmetry as helicity) observable~for the lower-order harmonics is 
associated with the topological properties of the material, the topological Chern number,~$C$~\cite{SilvaNatPho2019}.~{\color{black}From Figs.~\ref{fig1}b) and~\ref{fig1}f), we observe that the normalized CD is zero or close to zero for the trivial phase (see also Fig.~\ref{fig:fig4}a)) across all harmonics, but distinctly nonzero for the topological non-trivial phase of the Haldane model.~In fact, the ${\rm CD}$ is $-1.0$, see Fig.~\ref{fig:fig4}c)~and Table~\ref{tab1}, which matches the topological index $C=-1$.~This is further supported by Figs.~\ref{fig:fig4}a) and~\ref{fig:fig4}c), where a nonzero CD is observed in the ellipticity dependence for corotating HOs in the topological phase (see Fig.~\ref{fig:fig4}c)), in stark contrast to the trivial phase.~Clear differences are observed between the HOs produced by the topological trivial phase and the non-trivial phase for both: the helicity and circular dichroism.~Note that the difference is significantly larger for the CD than for the $h$ in these two topologically distinct phases.~Our results indicate that CD is a more suitable observable than harmonic helicity, as it is more directly related to the Berry curvature~\cite{TornowArxiv2026}.

\begin{table}[h!]
\centering
\caption{{\color{black}Helicity ($h^{(n)}$) and circular dichroism (${\rm CD}^{(n)}$, see tex for Eqs.) for corotating harmonic orders (\( 4^\text{th} \), \( 7^\text{th} \), and \( 10^\text{th} \)) in different topological phases of the Haldane Model and also the retrieval from Figs.~\ref{fig:fig4}.}}
\label{tab:cd_helicity}
\begin{tabular}{@{} l
                @{\hspace{1.5cm}} S[table-format=1.4]
                @{\hspace{1.5cm}} S[table-format=1.4]
                @{\hspace{1.5cm}} S[table-format=1.4] @{}}
\toprule
\textbf{Haldane} & {\( 4^\text{th} \)} & {\( 7^\text{th} \)} & {\( 10^\text{th} \)} \\
\midrule
\addlinespace[0.5em]
\multicolumn{4}{@{}l}{\textbf{Helicity}} \\
\hspace{1em}Trivial        & 0.50 &  0.38 &  0.00  \\
\hspace{1em}Non-trivial    & -1.00 &  -0.42 &  0.00 \\
\addlinespace[0.8em]
\multicolumn{4}{@{}l}{\textbf{Circular Dichroism}} \\
\hspace{1em}Trivial    & -0.10 &  -0.20 & -0.00  \\
\hspace{1em}Non-trivial   & -1.00 & -1.00 & -1.00  \\
\bottomrule
\label{tab1}
\end{tabular}
\end{table}

 \begin{figure}[h]
    \centering 
    \includegraphics[width=0.9\textwidth]{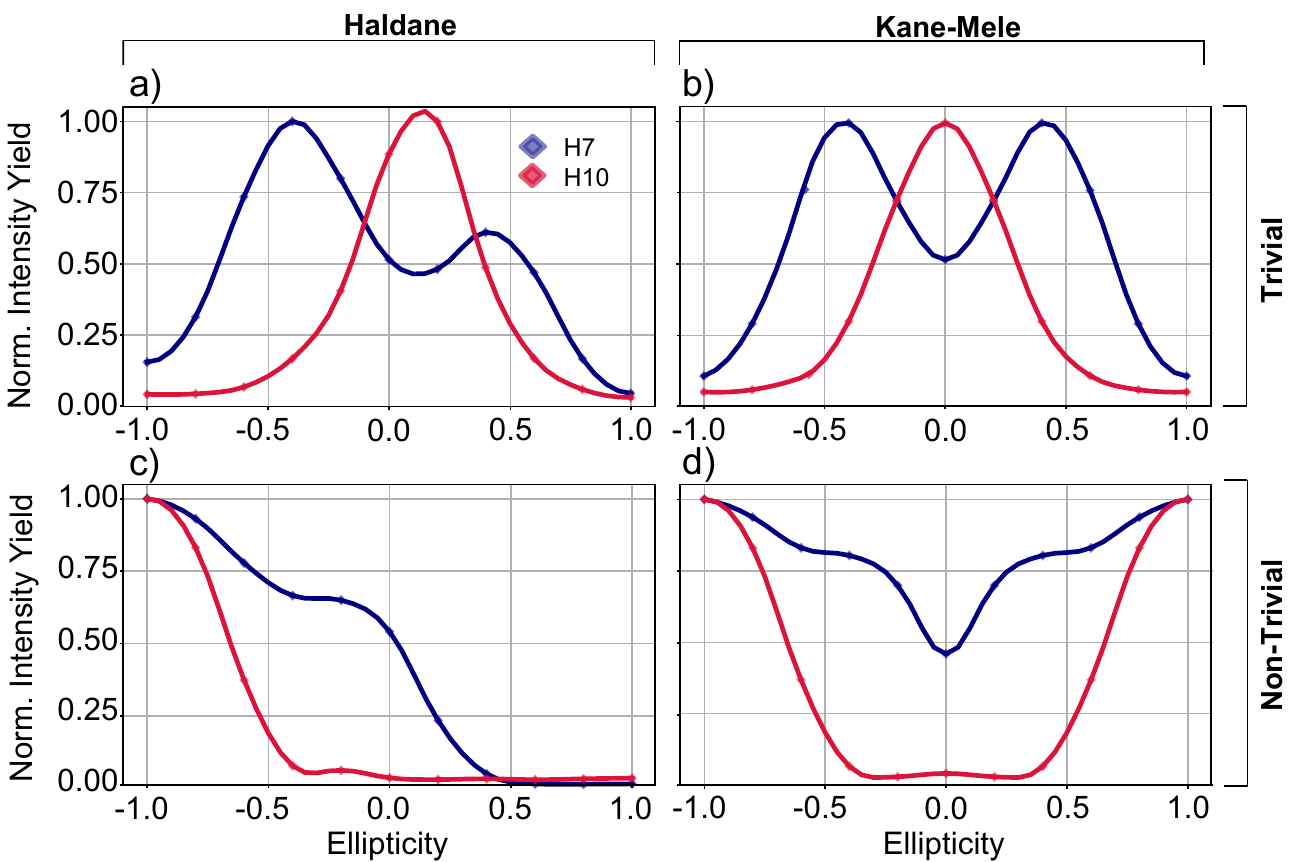}
    \caption{High harmonic spectroscopy in lower-order topological insulators (LOTIs): the ellipticity dependence. a) (topologically trivial) And c) (topologically non-trivial) show the normalized harmonic intensity yield as a function of the driving laser ellipticity for both corotating HOs: 7$^{\rm th}$ (blue line) and 10$^{\rm th}$  (red line), in case of 2D topological Chern insulators (CIs). Same in b) and d) but now for 2D TIs described by the Kane-Mele model. The laser parameters used in these simulations are the same that those employed in Fig.~\ref{fig1}. Note, in case of CIs there exists a strong asymmetry in the ellipticity dependence between left-handed circularly polarized (LCP) laser and right-handed circularly polarized (RCP) laser, namely, circular dichroism. In contrast, for TIs there is a symmetric behaviour of the intensity yield where the maximum intensity emissions are for LCP and RCP, but its minimum is about zero ellipticity, i.e.~linearly polarized driving laser. {\color{black}Discrete points correspond to the calculated values, and the solid curve is obtained by interpolating these points.}}
    \label{fig:fig4}
\end{figure}
\noindent {\color{black}In addition, we compute the helicity~and the circular dichroism (see Fig.~\ref{fig:fig4}b) and \ref{fig:fig4}d)) for the 2D TI described by the Kane--Mele model and found both to be negligible.~These observables do not capture the topological nature of TIs, where the QSHE is the underlying mechanism.~This is analogous to the detection of the IQHE, where the Hall {\it charge current} is the relevant observable, in contrast to the QSHE, where the {\it spin current} is the important quantity~\cite{KlitzingPRL1980,MarkusScience2007}.~Here, we draw an analogy: instead of helicity and circular dichroism, for TIs we measure a different observable based on high-harmonic generation, the harmonic intensity yield as a function of the driving laser ellipticity (see Fig.~\ref{fig:fig4}). The results reveal an {\color{black}anomalous ellipticity dependence (AED)} that cannot easily be explained by conventional HHG mechanisms in solids, where e--h annihilation typically occurs near the spatial birth location for LP driving fields, but is suppressed for circular polarization due to the lateral displacement of the e--h trajectory along a spiral path~\cite{GaardeOPG2022}.~This behavior suggests that {\it ellipticity dependence} is an 
observable for probing topological responses and phase transitions in TIs.~Moreover, our results for 2D TIs are in agreement with the experimental observations reported by Baykusheva {\it et al.}~\cite{BaykushevaNanoLetters2021}.~{\color{black}However, Ref.~\cite{NeufeldPRX2023} reported that this AED can also be found in topologically trivial materials, i.e., ``emission originating from 2D trivial bulk states''. Notably, the contribution from {\it topological edge states} was not considered in that theoretical study.~Therefore, the role of edge states remains an open question, {\color{black} by including edge and corner states in our theoretical model, we aim to clarify their role in the HHG response}.~Here, we introduce a complementary observable, namely, the relative intensity of the total harmonic emission from topological bulk and edge states compared to the trivial case.~As we have introduced and studied above in Figs.~\ref{fig:fig2} and \ref{fig:fig3}, in general, the HHG produced from the edge states has an intensity at least {\it two orders of magnitude larger} in comparison to the bulk emission for both models of our LOTI materials (all models have the same energy bulk bandgaps).~These results suggest that one should consider all possible channels for HHG emission from trivial and non-trivial topological materials, the radiation emitted from the topological bulk and edge states.}}

\noindent We now turn to the question of whether {\it higher-order topological information} can also be probed by ultrafast high-harmonic spectroscopy. To this end, we consider a breathing Kagome lattice structure and examine its HHG response in the next section.
}

\textit{higher-order topological insulators} 

\noindent In this section, we study the high-harmonic spectroscopy of higher-order topological insulators, focusing specifically on Kagome materials with a minimum of three-band structure (Nb$_3$TeCl$_7$, Nb$_3$Br$_8$, Nb$_3$Cl$_8$, etc.).~The Kagome lattice is a two-dimensional periodic arrangement of triangles sharing corners of atoms or sites A, B and C, where each pair of adjacent triangles is inverted with respect to the other, forming a pattern of hexagons interlaced with triangles, as illustrated in~Fig.~\ref{fig:fig0}.~Its Bravais lattice is hexagonal with a three-site basis, leading to a band structure that typically exhibits flat bands, Dirac cones, and van Hove singularities~\cite{GuoPRB2009,HuNatureCommun2022}.~These geometric features make the Kagome lattice an ideal platform to explore the interplay between geometry, electronic correlations, and topology~\cite{EzawaPRL2018,MarkusScience2007}.~In the presence of spin--orbit coupling, lattice distortions, or staggered hopping (such as in the breathing Kagome variant), the system can realize nontrivial topological phases, including Chern insulators~\cite{OhgushiPRB2000}, QSHE~\cite{MarkusScience2007,GuoPRB2009, BolensPRB2019}, and higher-order topological insulators~\cite{EzawaPRL2018,KempkesNatureMat2019}.~{\color{black}We focus on the Kagome model that possesses distinct topological or geometrical features as topological {\it bulk states}, topological {\it edge states} and topological {\it corner states} for trivial, non-trivial and  semimetallic phases~\cite{EzawaPRL2018,BolensPRB2019}.~{\color{black}Based on these features, we compute the full HHG spectra driven by different laser configurations. For the HOTI and trivial phases, the parameters were obtained from a Wannier-based tight-binding fit to the DFT band structure of Nb$_3$TeCl$_7$ (NTC), employed here as the reference system for the LOTI phases. The first-principles calculations were performed using VASP \cite{KressePRB1996,KressePRB1999}, and the low-energy bands were projected onto maximally localized Wannier functions using Wannier90 \cite{MostofiCPC2008,PizziJPCM2020}. For the semimetal phase, we employed parameters within the same energy range but corresponding to a semimetallic regime.} 
In the HOTI phase, the bulk and edge may remain insulating, while topological states (see Fig.~\ref{fig:fig0} Kagome lattice) are localized at the lattice {\it corners}, reflecting the bulk--boundary correspondence generalized to HOTI. {\color{black}Further details of the Kagome lattice structure and Hamiltonian model are provided in Appendix~\ref{AppendixA.3}.}
This combination of symmetry, geometry, and tunable topological phases makes the Kagome lattice a paradigmatic model for studying topological band theory.~{\color{black}Moreover, we employ the SSH model as a paradigm, as described in Refs.~\cite{EzawaPRL2018,KempkesNatureMat2019}. Both models share multiple similarities, since the SSH model can be viewed as a one-dimensional simplification of one of the edges of the triangular Kagome structure shown in Fig.~\ref{fig:B-E-C}. In this context, the edge states of the SSH model are analogous to the corner states of the Kagome HOTI, as both originate from the same topological mechanism associated with the alternation of hopping amplitudes.}

\noindent Here we investigate how the HHG can or not encode this topological information (topological {\it corner states}) by means of calculating the helicity, circular dichroism, ellipticity dependence and relative intensity emissions while a MIR laser drives the {\color{black}Kagome lattice}.~Our calculations are shown in~Fig.~\ref{fig:fig5}. 
\begin{figure}[h!]
    \begin{center}
    \includegraphics[width=0.8\textwidth]{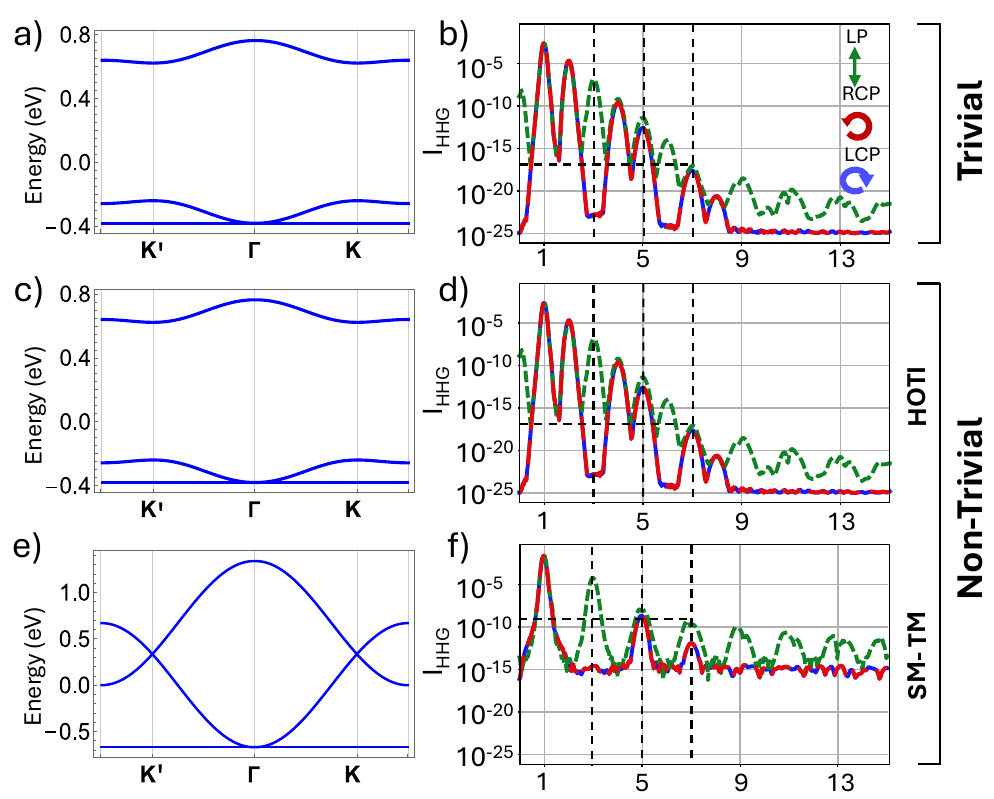}
    \end{center}
    \caption{{\color{black} High-harmonic generation from higher-order topological insulators, {\it \color{black}topological bulk state emissions}: Panels a) and b) show the band structure of our Kagome lattice model and the corresponding HHG spectra for the {\it trivial phase} under linearly polarized (LP, green dashed line), right-handed circularly polarized (RCP, red dashed line), and left-handed circularly polarized (LCP, blue solid line) driving lasers.~Panels c) and d) present the same analysis for the HOTI phase {\color{black}(non-trivial)}.~Panels e) and f) depict {\it \color{black}the energy band dispersions} and HHG spectra for the semimetal topological material (SM-TM). For the Kagome material, the lattice constant is \(a_0 \sim 7.0\, \text{\AA}\).~In the {\it topologically trivial phase}, the hopping parameters are \(t_a= 0.0123\)~a.u.~($\sim$ 0.33~eV) and \(t_b = 0.00173\)~a.u.~($\sim$ 0.047~eV) with ratio \(t_a/t_b\sim7.02\).~In the  HOTI {\it topologically non-trivial phase},  the parameters are reversed, such that \(t_b \rightarrow t_a\) and \(t_a \rightarrow t_b\)~where the ratio is \(t_a/t_b\sim 0.142\) (see Appendix \ref{AppendixA.3a})).~The hopping parameters also for the {\it SM-TM topologically non-trivial phase} are equal, with \(t_a = t_b = 0.0123\)~a.u.~($\sim$ 0.33~eV). The vertical black dashed lines connect the HOs:~$3^{\rm rd}$, $5^{\rm th}$ and $7^{\rm th}$ for the three different topological phases b), d) and f), respectively. Horizontal lines clear indicate that the HO $7^{\rm th}$ for the SM-TM has about three orders of magnitude larger that trivial and HOTI topological corner states, respectively.~Similarly to the LOTI case, the model parameters were fitted to preserve a band gap of approximately $0.8$~eV.}}
\label{fig:fig5}
\end{figure}
The trivial and non-trivial topological bands are exactly the same as depicted in Figs.~\ref{fig:fig5}a) and~\ref{fig:fig5}c), though the $P_3$ topological index is not null for the HOTI phase, it is null for trivial.~{\color{black}These topological {\it corner states} are included in the model used here to compute the full high-harmonic emission:~the calculated HHG spectra from topological bulk states, the topological edge states, and the topological corner states.}~We start then by showing the HHG spectra for {\color{black}trivial, HOTI and topological semimetal from the bulk states}.~{\color{black}The high harmonic spectra corresponding to} trivial and HOTI phases are presented in Figs.~\ref{fig:fig5}b) and \ref{fig:fig5}d), respectively.~The HHG spectra do not exhibit any clear difference between the harmonics produced by  LP, LCP and RCP driving lasers~{\color{black}for the parameters considered here in~Fig.~\ref{fig:fig5}}.~We also perform calculations of high-order harmonics by manipulating the location of the topological flat band at the bottom as a valence band and at the top as conduction one. Once again, the similar harmonic spectra were found for both trivial and non-trivial topological {\it bulk} Kagome lattice. 
\begin{figure}
    \centering
    \includegraphics[width=1\linewidth]{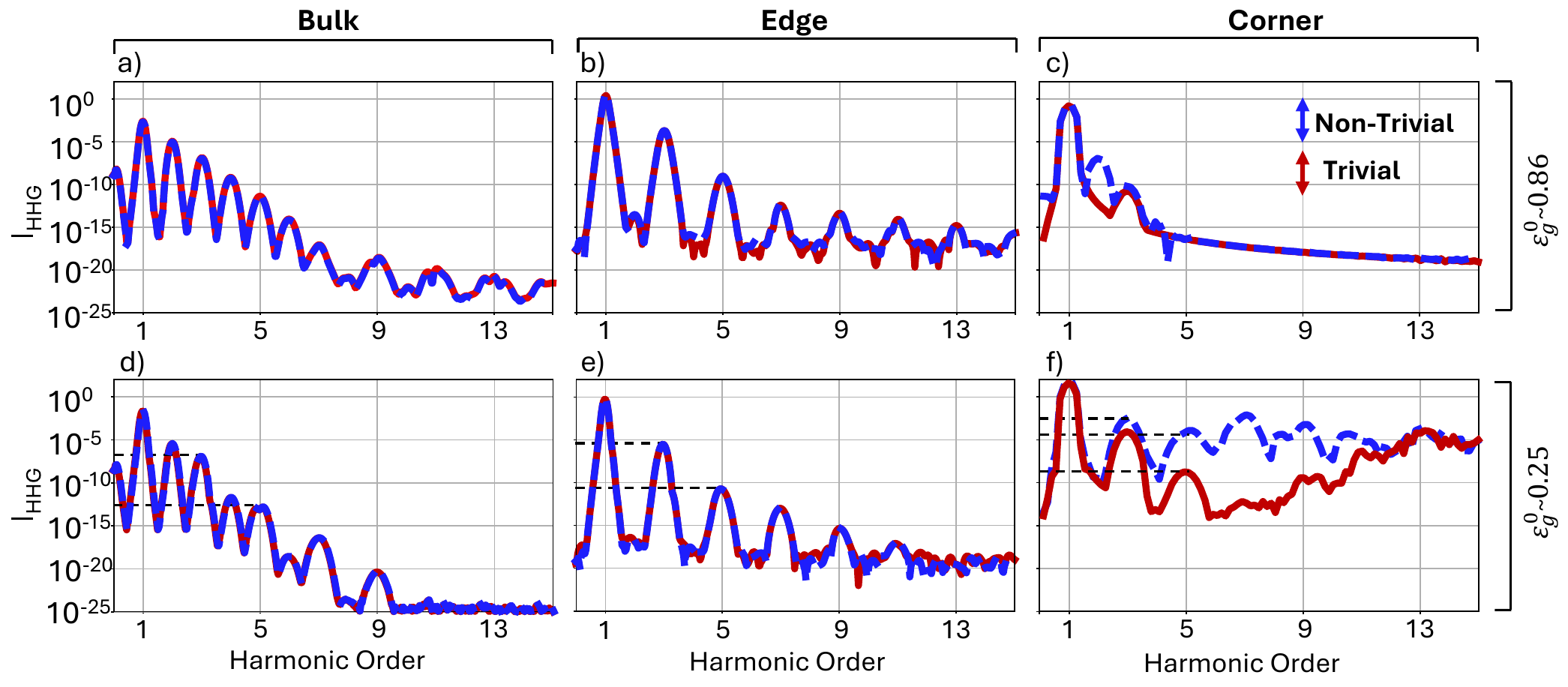}
    \caption{{\color{black}Kagome HOTIs: high harmonic emission from {\it bulk, edge,} and {\it corner topological states.} Panels a), b) and c) show the harmonic emission from the \textit{topological bulk states}, \textit{edge states}, and \textit{corner states}, respectively.~The energy gap is characterized by a bulk bandgap of $\varepsilon_g^{(0)} \sim 0.86$~eV.~The red (blue) dashed line denotes the trivial (non-trivial) phase.~The energy bands are shown in Fig.~\ref{fig:fig5}a) and \ref{fig:fig5}c).~For this case, the HOTI phase uses the hopping parameters $t_a = 0.00173$~a.u.~($\sim 0.047$~eV) and $t_b = 0.0123$~a.u.~($\sim 0.33$~eV), while in the trivial phase these values are interchanged ($t_b \rightarrow t_a$, $t_a \rightarrow t_b$).~c)~The corresponding HHG spectra from SSH model were performed using $\delta = \pm 0.49$~a.u.~for the topologically trivial and non-trivial phases.~Panels d) and e) display analogous HHG spectra for a system with a smaller bulk bandgap of $\varepsilon_g^{(0)}~\sim 0.25$~eV.~The bands are shown in Fig.~\ref{fig:FigS4}.~Here, the HOTI phase employs $t_a = 0.0037$~a.u.~($\sim 0.10$~eV) and $t_b = 0.0067$~a.u.~($\sim 0.183$~eV), with the trivial phase again obtained by inverting these values.~f)~The SSH model parameters for this case are $\delta = \pm 0.15$~a.u.~As before, harmonic spectra are shown for bulk, edge, and corner states. In particular, the smaller-gap configuration highlights a more pronounced contrast between the trivial and topological phases at the corner level.~In the edge geometry, translational symmetry is broken along the $y$-axis, while the corner geometry corresponds to a finite system described within the SSH model framework. All simulations use the same laser parameters as those in Fig.~\ref{fig1}.}}    

    \label{fig:fig6}
\end{figure}

\noindent{\color{black}Since the topological bulk emission does not show a clear signature of topology in this Kagome lattice throughout the HHG process, we naturally turn to the contributions from {\it boundary states}.~We therefore examine how the emissions from the topological {\it edge states} and topological {\it corner states} affect the harmonic response when included in the calculation for HOTIs.~To this end, we follow the same procedure as described in the above section for lower-order topological insulators.~The results are shown in Fig.~\ref{fig:fig6}~for the HHG emitted from the topological~{\it bulk states}, {\it edge states} and {\it corner states} corresponding to Figs.~\ref{fig:fig6}a), \ref{fig:fig6}b) and \ref{fig:fig6}c), respectively.~We observe that for the emitted topological {\it bulk states} and {\it edge states} there are not a {\it significant} difference between the trivial and non-trivial topological phases.~Nevertheless, the emitted harmonics by the topological {\it corner states} show differences in the even harmonics. But, these topological corner states do not represent a conventional HHG spectrum, since the cut-off is very low (cut-off $\leq $ HO5$^{\rm rd}$).~Hence, we cannot conclude whether the emission from a system with $\delta = -0.49$~a.u.~carries corner-related topological information within the studied region.

\noindent In order to prove that the existence of~{\it topological corner states} can be encoded in the HHG spectrum, we carried out additional theoretical studies by adjusting the \(\delta\) parameter of the SSH model, using a small band gap, in order to obtain a system that exhibits a larger cut-off, i.e.,~comparable to the cut-offs of the topological bulk and topological edge states.~These calculations are shown in Figs.~\ref{fig:fig6}d), \ref{fig:fig6}e), and \ref{fig:fig6}f).~Indeed, the HHG emission pattern observed in Figs.~\ref{fig:fig6}a)~and~\ref{fig:fig6}b) is also repeated in the new topological system shown in Figs.~\ref{fig:fig6}d) and~\ref{fig:fig6}e).~But, and interestingly, in the emitted harmonics from~{\it topological corner states} under the new or current material parameters, we find an {\it enhancement} of one to two orders of magnitude (see Fig.~\ref{fig:fig6}f)) in comparison with the {\it topological bulk {\rm or} edge states}. This {\it enhancement} distinguishes the HHG response of the topological Kagome  lattice from that obtained using the previous material parameters (see Figs.~\ref{fig:fig6}a)~and~\ref{fig:fig6}b)). {\color{black}Crucially, this enhancement can be regarded as a characteristic signature of topological corner states, so that comparing their yield with that of a trivial material provides a direct diagnostic of the underlying topological phase.} 

\begin{figure}
    \centering
    \includegraphics[width=0.8\linewidth]{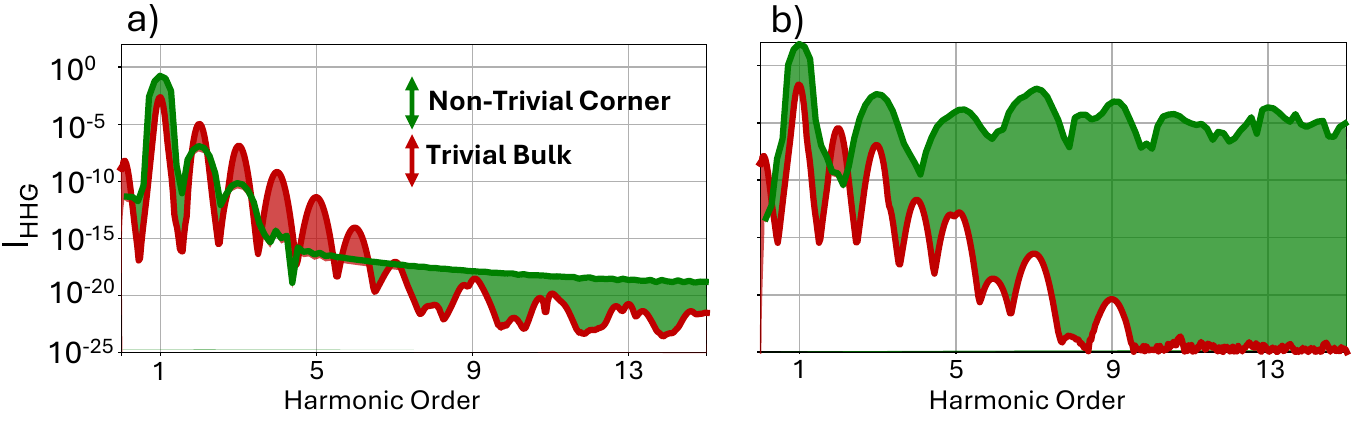}
\caption{{\color{black}High-harmonic generation spectra for the trivial and topologically non-trivial HOTI phases in Kagome lattices. Panels~a) and b) correspond to the two parameter sets used in Fig.~\ref{fig:fig5}.~In each panel, the emission from the {\it trivial bulk states} (red line) is compared with the emission from the {\it non-trivial corner states} (green line).~The topological HHG emission from corner {\it states} exhibits a different behavior compared to the trivial bulk emission and depends on the topological corner parameters.~Since the cutoff of the trivial phase is better reproduced in b), we believe that these results suggest that HHS suitably captures topology through relative intensity enhancement (see Ref.~\cite{BauerPRL2018, BauerPRB2019}).}}
    \label{fig:fig7}
\end{figure}

\noindent {\color{black}Similarly to the LOTI case, we compare the HHG spectra generated from the bulk of the material with those obtained when topological corner states are taken into account.~The comparison is performed for two sets of parameters shown in Fig.~\ref{fig:fig7}.~For the first parameter set (Fig.~\ref{fig:fig7}(a)), the bulk emission is found to be stronger than that originating from the topological corner states of the finite system.~In contrast, for the second parameter set, the {\it topological corner states} exhibit a significantly higher emission intensity than the bulk states.~Here, we show that additional system parameters, such as details of the band structure, play a crucial role in determining the relative intensity, specially for {\it corner} states.~The HHG spectra results shown here may still be reflected in the observed emission trends, which are intrinsic to systems hosting {\it edge states} (LOTIs) and {\it corner states} (HOTIs). In this sense, comparing the harmonic emission from {\it bulk states} and {\it corner states} can provide a qualitative indication of the underlying topology in Kagome lattice systems or, more generally, HOTI structures (Fig.~\ref{fig:fig7}).~However, in regimes where the topological states are suppressed or not well defined, the emission from {\it corner states} can no longer be meaningfully distinguished from other contributions, such as {\it bulk} or {\it edge states}.}~Moreover, we investigate how the HHG spectra from the {\it topological corner states} evolve as a function of the parameter $\delta$ in the SSH model, and we present these results in the Appendix~\ref{AppendixB.3}. There, our results show the behavior of the emissions varies as a function of $\delta$, which controls the band gap of the material, and how the increase in the harmonic intensity depends on this parameter, 
causing the topological signatures we observe to shift across the frequency range.

\noindent Further calculations of helicity and circular dichroism were performed and those quantities do not exhibit any different between trivial and HOTI phases.~In contrast, the semimetal topological material (SM-TM) shows minor differences between the harmonics generated by LCP and RCP laser pulses. Note, however, a relative {\it enhancement} about {\it one or two orders of magnitude} is observed for harmonics produced by LP lasers in comparison with the observed in trivial and HOTI phases.{\color{black}~This observation suggests that the HHG spectrum exhibits an {\it indirect} sensitivity to the topological properties of the Kagome lattice semimetal.}

\noindent To further investigate whether the {\color{black}topological information of {\it corner states} can be inferred from the {\it bulk states} via {\it the ellipticity dependence}, we compute the harmonic order yield as a function of the driving-laser ellipticity, as described in the LOTI section.}~The results are shown in Fig.~\ref{fig:fig8}.~Once again, we do not observe any unusual behavior between the topologically trivial and non-trivial phases for the Kagome models (see Figs.~\ref{fig:fig8}a),~\ref{fig:fig8}b), and~\ref{fig:fig8}c)), as was also observed for the TI in the previous section. In case of SM-TM, the harmonics likewise do not exhibit any atypical behavior.


\noindent Naturally, one may wonder whether this high-harmonic spectroscopy analysis demonstrates that the HHG mechanism can capture key topological information.~Based on the theoretical results presented here, we conclude that high-harmonic spectroscopy is a reliable observable for encoding topological information from the topological {\it edge states} and topological {\it corner states} in the cases of LOTIs and HOTIs, respectively.~Additional models may be explored to further disentangle the physics of Kagome materials; for instance, topological Kagome Chern insulators~\cite{OhgushiPRB2000}, in which a magnetic flux restores the topological features of the Haldane model. However, such studies lie beyond the scope of the present work.}\\


\begin{figure}[h!]
    \begin{center}
    \includegraphics[width=0.9\textwidth]{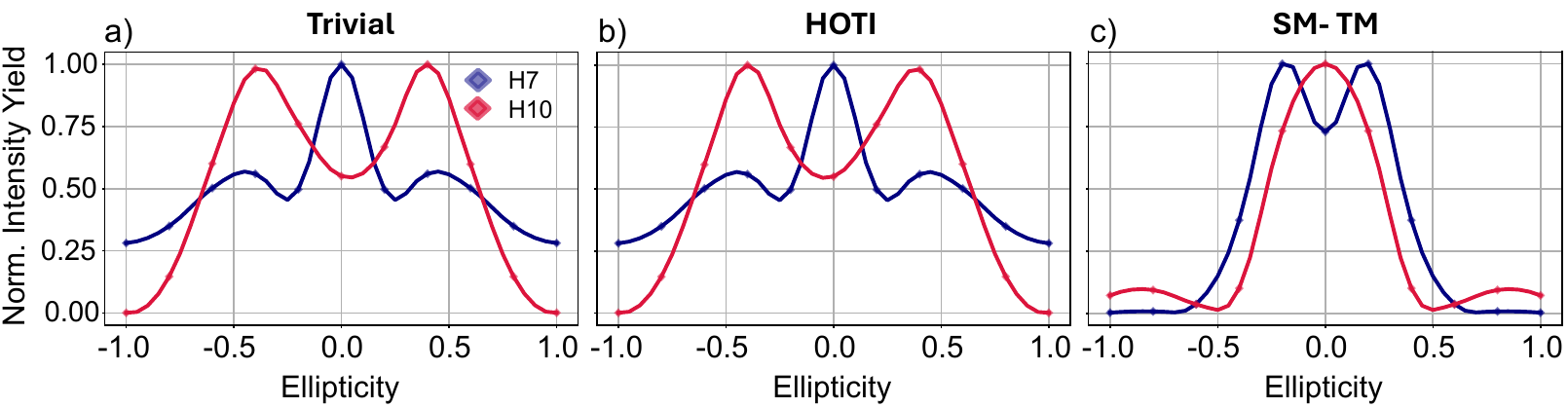}
    \end{center}
    \caption{High-harmonic spectroscopy in higher-order topological insulator (HOTI): ellipticity dependence.~Panels a), b), and c) show the ellipticity dependence of the normalized harmonic intensity as a function of the driving laser ellipticity for the trivial, HOTI, and semimetal phases, respectively.~Note that, for these Kagome structures, the trivial and HOTI phases show the same tendency for both the 7$^{\rm th}$ and 10$^{\rm th}$ HOs. The laser and material parameters used here are the same as those in Fig.~\ref{fig:fig5}.~{\color{black}The solid curves are obtained by interpolating the discrete data points.}}
    \label{fig:fig8}
\end{figure}

\textit{Conclusions} 

\noindent We studied high-harmonic spectroscopy based on high-order harmonic generation (HHG) in lower-order topological insulators (LOTIs) and higher-order topological insulators (HOTIs) by analyzing the helicity, circular dichroism, ellipticity dependence and {\color{black}the relative intensity emission produced by mid-infrared lasers.}~{\color{black}For LOTIs, we found that the high harmonics produced in the Haldane and Kane--Mele models exhibit distinct characteristics, suggesting that the all nonlinear optical responses or high harmonic orders can prove or diagnose, 
topological phases} and transitions {\color{black} in Chern insulators and topological insulators}.~For Chern insulators, all of these experimental observables {\color{black}(helicity, circular dichroism, anomalous ellipticity dependence, and relative intensity emission yield)}~display features indicative of {\color{black}topology, including the {\it topological edge states} by means of a clear {\it intensity enhancement} of the emitted harmonics from topologically {\it non-trivial states} in comparison to {\it topologically trivial}.}~In case of two-dimensional topological insulator described by the Kane--Mele model, the harmonic intensity yield as a function of ellipticity shows an atypical behavior, namely, anomalous ellipticity dependence, which agrees with recent experimental observations from Bi$_2$Se$_3$~\cite{BaykushevaNanoLetters2021,HeideNaturePhotonics2022}.~In addition, we find an enhancement of the harmonic intensity yield for the topologically nontrivial edge states compared with the trivial phase similar to the case of the Haldane model.

\noindent{\color{black}For HOTIs, in particular the topological Kagome lattices,~we found numerical evidences suggesting that the {\it intensity yield} produced by {\it topological corner states} has a relative {\it enhancement} of one or two-orders of magnitude in comparison with the {\it topological bulk states}.~This clearly suggests  that the {\it topological corner states} are 
encoded in the {\it intensity yield} of the harmonic emission from the HOTI phase and that they play an intrinsic role in the HHG mechanism.}~{\color{black}Note, however, that there exist HOTI material conditions under which the HHG spectra emitted from the topological Kagome lattice are challenging to characterize (see the HHG from HOTI section). In such cases, the observables, helicity, circular dichroism, ellipticity dependence, and intensity analysis, cannot be directly associated with topological information.~However, when the system exhibits a topological semimetal phase, we observe a relative enhancement of the harmonic yields from the Kagome structure by more than three orders of magnitude.~This {\it enhancement} of the HHG spectra, together with the contribution from the topological {\it corner states}, constitutes an approximated signature of HOTIs in the nonlinear optical response.~Finally, our calculations show that the {\it topological edge states} (in the case of LOTI materials) and the {\it topological corner states} (in the case of HOTI materials) are the channels that most effectively capture the topological information in the HHG mechanism.~This pattern is consistent across both LOTI and HOTI topological classifications.~Our results further suggest that the characterization of topological materials via HHS must account for all emission channels topological bulk, edge, and corner states.~This work paves the way for future efforts to {\it directly} link topological invariants (or related to the invariants such as Berry phases and Berry curvatures) with HHS observables, a direction we are currently pursuing.} }  \\

\textit{Acknowledgment}

B.L., C.B., J.M., and A.C. acknowledge the Sistema Nacional de Investigaci\'on de Panam\'a for partial financial support. We thank~Prof.~M.~Santamar\'ia and Prof.~N.~Correa for logistical support at the Centro de Investigaci\'on con T\'ecnicas Nucleares, Universidad de Panam\'a.~A.C. thanks the Sherlock Cluster at Stanford University for generous computational time. 
This work was supported by the Institute for Basic Science (IBS), Korea under Project Code IBS-R014-A1. D.K. acknowledges support from the National Research Foundation of Korea grants (Grant no. NRF-2023R1A2C2007998). This study was also supported by the MSIT (Ministry of Science and ICT), Korea, under the ITRC (Information Technology Research Center) support program (Grant no. IITP-2023-RS-2022-00164799) supervised by the IITP (Institute for Information Communications Technology Planning Evaluation).
W.Gao thanks finicial support from National Natural Science Foundation of China (Grant number: 12474309).
M.J. gratefully acknowledges funding by the Deutsche Forschungsgemeinschaft (DFG, German Research Foundation) through IRTG 2676/1 ‘Imaging of Quantum Systems’, project no. 437567992. S.G. acknowledges the support from US Department of Energy, Office of Science, Basic Energy Sciences, Chemical Sciences, Geosciences, and Biosciences Division through the AMOS program.

\newpage

\appendix
\titlespacing*{\subsection}
  {0pt}{1.5ex}{0.5ex }
\titlespacing*{\section}
  {0pt}{1.5ex}{0.5ex }
\titlespacing*{\subsubsection}
  {0pt}{1.5ex}{0.5ex }

\normalsize%
\setlength\abovedisplayskip{1pt}%
\setlength\belowdisplayskip{8pt}%
\setlength\abovedisplayshortskip{-8pt}%
\setlength\belowdisplayshortskip{4pt}%

\section{Tight Binding Approximation Models}\label{AppendixA}
\renewcommand{\theequation}{\thesection\arabic{equation}}
\setcounter{equation}{0} 

\renewcommand{\thefigure}{\thesection.\arabic{figure}}
\setcounter{figure}{0}
This appendix includes three target models used to describe the physical picture and high-order harmonic generation: the Haldane and Kane--Mele models for distinguishing lower-order topological insulators, and a simple breathing Kagome lattice tight-binding approximation for higher-order topological insulators. 
\subsection{Haldane Honeycomb lattice}\label{AppendixA.1}
\subsubsection{Bulk Hamiltonian}\label{AppendixA.1a}
\begin{figure}[h!]
    \centering
    \includegraphics[width=0.6\linewidth]{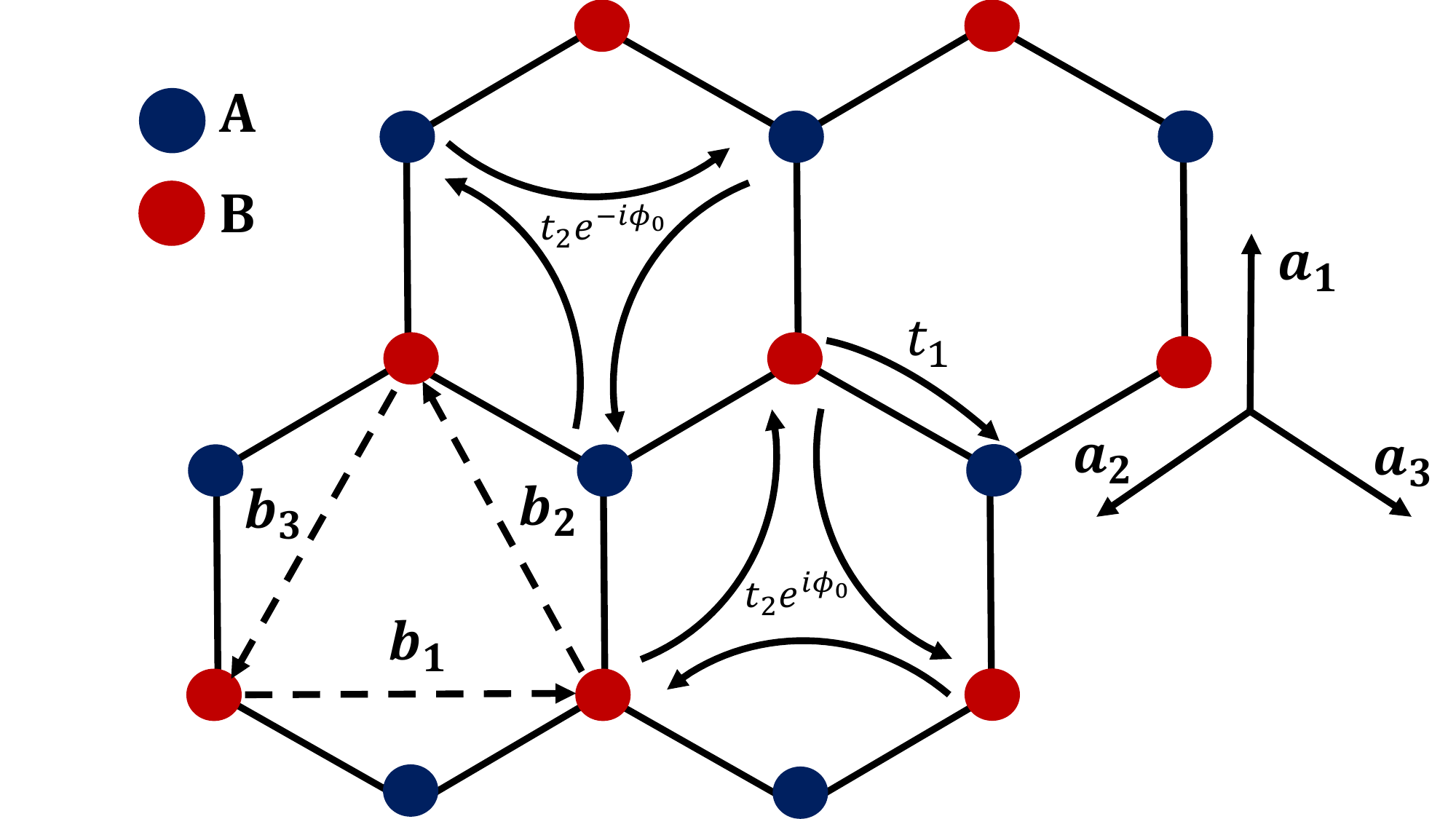}
    \caption{Honeycomb crystalline lattice. Fundamental structure underlying the Haldane and Kane--Mele models. Red and blue dots represent the two sublattices, while black lines indicate the nearest-neighbor (NN) connections. These NN hoppings are represented by \(t_1\), describing hopping from an A site to a B site. The lines linking across each hexagon correspond to the next-nearest-neighbor (NNN) hoppings, represented by \(t_2 e^{\pm i\phi_0}\), which also encode the fluxes within the hexagons. The lattice is further defined by the basis vectors \(\mathbf{a}_i\).%
}
    \label{fig:FigS1}
\end{figure}

\noindent The Haldane model is a fundamental two-band tight-binding system on a honeycomb lattice, which can realize either a topologically trivial or a Chern insulating phase depending on its parameters~\cite{HaldanePRL1988}. The honeycomb lattice consists of two sublattices A and B. The nearest neighbor (NN) hopping vectors \(\mathbf{a}_i\) connect sites from sublattice A to sublattice B, while the next nearest neighbor (NNN) hopping vectors \(\mathbf{b}_i\) connect sites within the same sublattice (i.e., \({\rm A} \to {\rm A}\) or \({\rm B} \to {\rm B}\)). These vectors are defined as:
\begin{align}
\mathbf{a}_1 &= \left(0, 1\right) a_0, \quad
\mathbf{a}_2 = \frac{1}{2}\left(-\sqrt{3}, -1\right) a_0, \quad
\mathbf{a}_3 = \frac{1}{2}\left(\sqrt{3}, -1\right) a_0, \\
\mathbf{b}_1 &= \left(\sqrt{3}, 0\right) a_0, \quad
\mathbf{b}_2 = \frac{1}{2}\left(-\sqrt{3}, 3\right) a_0, \quad
\mathbf{b}_3 = \frac{1}{2}\left(-\sqrt{3}, -3\right) a_0,
\end{align}
where \(a_0\) is the lattice constant (see Fig.~\ref{fig:FigS1}). The Hamiltonian in momentum space can be expressed as a \(2 \times 2\) matrix,
\begin{equation}
H_1(\mathbf{k}) = B_{0,\mathbf{k}} \sigma_0 + \mathbf{B}_{\mathbf{k}} \cdot \boldsymbol{\sigma},
\label{eq:H1}
\end{equation}
where \(\sigma_0\) is the identity matrix, and \(\boldsymbol{\sigma} = (\sigma_1, \sigma_2, \sigma_3)\) are the Pauli matrices. The functions \(B_{0,\mathbf{k}}\) and \(\mathbf{B}_{\mathbf{k}} = (B_{1,\mathbf{k}}, B_{2,\mathbf{k}}, B_{3,\mathbf{k}})\) encode the hopping and on-site terms,and are given by
\begin{align}
B_{0,\mathbf{k}} &= 2t_2 \cos(\phi_0) \sum_{i=1}^3 \cos(\mathbf{k} \cdot \mathbf{b}_i) \\
B_{1,\mathbf{k}} &= t_1 \sum_{i=1}^3 \cos(\mathbf{k} \cdot \mathbf{a}_i), \\
B_{2,\mathbf{k}} &= t_1 \sum_{i=1}^3 \sin(\mathbf{k} \cdot \mathbf{a}_i), \\
B_{3,\mathbf{k}} &= M_0 - 2 t_2 \sin(\phi_0) \sum_{i=1}^3 \sin(\mathbf{k} \cdot \mathbf{b}_i).
\end{align}
Here, \(t_1\) and \(t_2\) denote the NN and NNN hopping amplitudes, respectively; \(M_0\) is an on-site energy term that breaks inversion symmetry (IS); and \(\phi_0\) is a phase related to a local magnetic flux, which breaks time-reversal symmetry (TRS). Diagonalizing the Hamiltonian Eq.~\ref{eq:H1} yields the energy bands
\begin{equation}
\varepsilon_{\pm}(\mathbf{k}) = B_{0,\mathbf{k}} \pm |\mathbf{B}_{\mathbf{k}}|,
\end{equation}
The eigenstates in the Bloch form are
\begin{equation}
|\Phi_{\pm,\mathbf{k}}\rangle = e^{i \mathbf{k} \cdot \mathbf{r}} |u_{\pm, \mathbf{k}}\rangle,
\end{equation}
with the lattice-periodic spinors
\begin{equation}
\begin{aligned}
    |u_{+,\mathbf{k}}\rangle &= \frac{1}{N_{+,\mathbf{k}}}
    \begin{pmatrix}
        B_{3,\mathbf{k}} + |\mathbf{B}_{\mathbf{k}}| \\
        B_{1,\mathbf{k}} + i B_{2,\mathbf{k}}
    \end{pmatrix}, \\[6pt]
    |u_{-,\mathbf{k}}\rangle &= \frac{1}{N_{-,\mathbf{k}}}
    \begin{pmatrix}
        i B_{2,\mathbf{k}} - B_{1,\mathbf{k}} \\
        B_{3,\mathbf{k}} + |\mathbf{B}_{\mathbf{k}}|
    \end{pmatrix}.
\end{aligned}
\label{eq:eigenstates}
\end{equation}
\noindent where the \((+)\) sign corresponds to the conduction band and the \((-)\) sign to the valence band, with ${N_{\pm,\mathbf{k}}}$, a suitable normalization constant.
The topological character of the Haldane model is captured by the Chern number \(C_m\), which acts as a topological invariant for the \(n\)-th energy band. It is defined as the 2D integral over the first Brillouin zone (BZ) of the Berry curvature \(\mathbf{\Omega}_m(\mathbf{k})\):
\begin{equation}
C_m = \frac{1}{2\pi} \int_{\mathrm{BZ}} d^2 \mathbf{k} \,\cdot \mathbf{\Omega}_m(\mathbf{k}).
\end{equation}
\noindent The Berry curvature can be computed from the lattice-periodic eigenstates \(|u_{m,\mathbf{k}}\rangle\) and their derivatives with respect to \(\mathbf{k}\). More specifically, the diagonal elements of the dipole transition matrix, 
\begin{equation}
\mathbf{\xi}_m(\mathbf{k}) = i \langle u_{m,\mathbf{k}} | \nabla_{\mathbf{k}} | u_{m,\mathbf{k}} \rangle,
\end{equation}
define the Berry connection of the band, which encodes the parallel transport of the wavefunction phase in momentum space. The Berry curvature is then obtained as the gauge-invariant curl of this connection,
\begin{equation}
\mathbf{\Omega}_m(\mathbf{k}) = \nabla_{\mathbf{k}} \times \mathbf{\xi}_m(\mathbf{k}).
\end{equation}
\noindent When \(C_m = 0\), the system is topologically trivial, whereas a nonzero Chern number \(C_m \neq 0\) indicates a Chern topological insulating phase, characterized by protected edge states and singular behavior in the off-diagonal dipole elements.
{\color{black}
\subsubsection{Edge Hamiltonian}\label{AppendixA.1b}
}
{\color{black}
\noindent With the purpose of studying the emission from topological edge states, we briefly present here the effective Hamiltonian for the Haldane model. The model used consists of a spinless zigzag nanoribbon of honeycombs, assuming zero net magnetic flux within the unit cell. This model is presented in Ref.~\cite{TraversonpjQuatumMat2024}.
\noindent We consider a finite-width hexagonal ribbon composed of $N_x$ unit cells along the periodic $x$-direction and $N_y$ unit cells along the finite $y$-direction. Periodic boundary conditions are applied along $x$, while open boundary conditions along $y$ generate the characteristic zigzag chains.~The Hamiltonian includes NN and NNN hoppings similar to those used in the bulk model,
}
\begin{equation}
    H_0 = m \sum_{ i}\epsilon_i \, c_i^\dagger c_i + t_1 \sum_{\langle i,j\rangle} c_i^\dagger c_j  +t_2 \sum_{\langle\!\langle i,j \rangle\!\rangle}  e^{-i\phi_{ij}} c_i^\dagger c_j  + \text{h.c.}
    \label{eq_H:H_0}
\end{equation}
{\color{black}In the following, we replace the index $i$ with $(m, n, s)$, which denotes the lattice site, where $(m, n)$ describes the unit cell and $s$ labels the A and B sites within that unit cell. Because the ribbon is periodic only along \(x\), we perform a partial Bloch transformation
}
\begin{equation}
    c_{m,n,s} = \frac{1}{\sqrt{N_k}} 
    \sum_{k_x} e^{i k_x x_{mns}} c_{k_x,n,s}
    \label{eq_H:TF_partial}
\end{equation}

{\color{black}where \((x_{mns},\, y_{mns})\) represents the coordinate of site \(s\) in the unit cell.~The Hamiltonian of the nanoribbon becomes block--tridiagonal, which yields a $k_x$-dependent Hamiltonian matrix.
}
\begin{equation}
    H(k_x) \in \mathbb{C}^{N_s \times N_s}.
    \label{eq_H:H_k}
\end{equation}

{\color{black}
The diagonal blocks contain the on-site mass term $m$ together with the next-nearest-neighbor
contributions, which in our ribbon geometry take the form
}
\begin{equation}
    h_{i,i} \;\rightarrow\; 
    \Big( 
         2 t_2 \cos(k_x  a+\phi) + m,\;
         2 t_2 \cos(k_x  a-\phi) - m
    \Big),
    \label{eq_H:Diagonal}
\end{equation}
{\color{black}while the first superdiagonal blocks correspond to the nearest-neighbor hoppings between adjacent
unit cells,}
\begin{equation}
    h_{i,i+1} \;\rightarrow\; 
    \Big(
        2t_1 \cos\!\big(k_x \tfrac{a}{2}\big),\;
        t_1
    \Big),
    \label{eq_H:1st_diagonal}
\end{equation}
{\color{black}and the second superdiagonal blocks encode the next-nearest-neighbor hoppings across two unit cells,
}
\begin{equation}
    h_{i,i+2} \;\rightarrow\; 
    \Big(
        2 t_2 \cos\!\big(k_x\tfrac{a}{2}-\phi\big),\;
        2 t_2 \cos\!\big( k_x\tfrac{a}{2}+\phi\big)
    \Big).
    \label{eq_H:2nd_diagonal}
\end{equation}

{\color{black}Finally, the lower diagonal blocks are given by the Hermitian conjugates of the corresponding
upper blocks,}
\begin{equation}   
    h_{i,i-1} = h_{i-1,i}^\dagger, 
    \qquad
    h_{i,i-2} = h_{i-2,i}^\dagger.
    \label{eq_H:h.c.}
\end{equation}

{\color{black} Explicitly, the Hamiltonian matrix takes the form}
\begin{equation}
\resizebox{\textwidth}{!}{$
H(k_x)=
\begin{pmatrix}
2 t_2 \cos(k_x a+\phi) + m 
    & 2t_1 \cos(k_x a/2) 
    & 2 t_2 \cos(k_x a/2-\phi) 
    & 0 & 0 & 0 & \cdots \\
h_{1,2}^\dagger 
    & 2 t_2 \cos(k_x a-\phi) - m 
    & t_1 
    & 2 t_2 \cos(k_x a/2+\phi) 
    & 0 & 0 & \cdots \\
h_{1,3}^\dagger 
    & h_{2,3}^\dagger 
    & 2 t_2 \cos(k_x a+\phi) + m 
    & 2t_1 \cos(k_x a/2) 
    & 2 t_2 \cos(k_x a/2-\phi) 
    & 0 & \cdots \\
0 
    & h_{2,4}^\dagger 
    & h_{3,4}^\dagger 
    & 2 t_2 \cos(k_x a-\phi) - m 
    & t_1 
    & 2 t_2 \cos(k_x a/2+\phi) 
    & \cdots \\
\vdots & \ddots & \ddots & \ddots & \ddots & \ddots & \vdots
\end{pmatrix}
$}
\end{equation}
{\color{black}
The off-diagonal blocks $h_{i,i+1}$ y $h_{i,i+2}$ are filled according to Eqs~\ref{eq_H:Diagonal} --~\ref{eq_H:h.c.}.
This procedure specifies the Haldane Hamiltonian for an arbitrary number $N$ of cells 
along the open direction.}

\subsection{Kane-Mele model}\label{AppendixA.2}

The Kane–Mele model describes a two-dimensional quantum spin Hall insulator (or topological insulator) that is protected by time-reversal symmetry (TRS) and characterized by the \(\mathbb{Z}_2\) topological invariant~\cite{Kane1PRL2005,Kane2PRL2005}. It can be understood as an extension of the Haldane model, formulated as a tight-binding Hamiltonian on the honeycomb lattice. The model includes nearest-neighbor (NN) hopping \(t_1\), next-nearest-neighbor (NNN) hopping terms \(t_2 \cos\phi_0\) (spin-independent) and \(t_2 \sin\phi_0\) (from intrinsic spin–orbit coupling), an on-site potential difference \(M_0\), and a Rashba term \(t_R\) due to a perpendicular electric field.  

\noindent Neglecting the Rashba term, spin-flipping is absent and \(S_z\) is a good quantum number. The Hamiltonian can then be written in momentum space as a block-diagonal \(4 \times 4\) matrix composed of two \(2 \times 2\) Haldane-model Hamiltonians:
\begin{equation}
H_{\text{KM}}(\mathbf{k}) =
\begin{pmatrix}
H_1(\mathbf{k}) & 0 \\
0 & H_1^*(-\mathbf{k})
\end{pmatrix},
\end{equation}
where \(H_1(\mathbf{k})\) corresponds to the spin-up Haldane model and \(H_1^*(-\mathbf{k})\) to the spin-down counterpart with effective flux \(\phi_0 \to -\phi_0\). This representation highlights the block structure of the Hamiltonian, emphasizing that the Kane--Mele model consists of two Haldane models coupled by time-reversal symmetry, where each spin sector acts independently.

\noindent The block can also be written explicitly in terms of Pauli matrices as Eq.~\ref{eq:H1}
with \(H_1\) (\(H_1^*\)) corresponding to spin \(\uparrow\) (\(\downarrow\)). This form is convenient for explicit calculations of eigenvalues, eigenstates, and topological quantities, such as the Berry curvature and spin Chern number. The topological character of the Kane–Mele model is captured by the spin Chern number
\begin{equation}
C_\sigma = \frac{1}{2} (C_+ - C_-),
\end{equation}
where \(C_\pm\) are the Chern numbers of the spin-resolved bands. The \(\mathbb{Z}_2\) invariant is then obtained as
\begin{equation}
\nu = C_\sigma \bmod 2.
\end{equation}
For the Haldane model, \(C = 0, \pm 1\), so the spin Chern number can take values \(C_\sigma = 0, \pm 1\). Nonzero spin Chern numbers \(C_\sigma = \pm 1\) correspond to the same topological class \(\nu = 1\), which defines the quantum spin Hall phase.

{\color{black}\noindent The edge Hamiltonian for the Kane--Mele model is constructed following the same zigzag nanoribbon approach employed for the Haldane model, including periodic boundary conditions along the \(x\)-direction and open boundary conditions along the transverse direction.}

\subsection{Kagome Model}\label{AppendixA.3}

{\color{black}
To study the contribution of the topological states, it is essential to open the boundary conditions (OBC), since these boundaries are the regions where the topological modes reside. If the surface, edge, or corner does not physically exist, it becomes impossible for such topological states to manifest in the system.
}
\begin{figure}[h!]
    \centering 
    \includegraphics[width=0.95\textwidth]{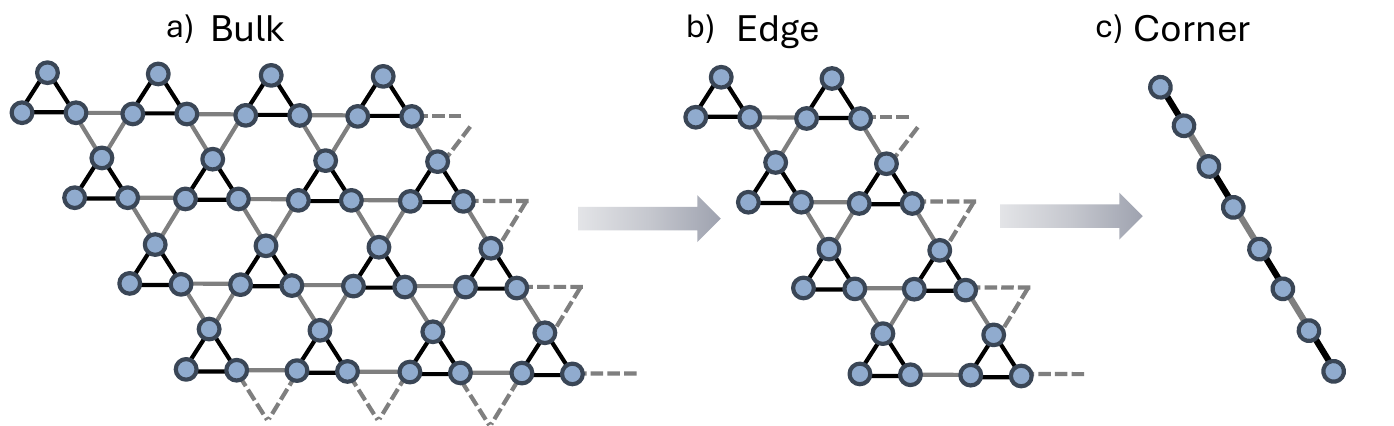}
    \caption{{\color{black}a) Shows the Kagome bulk structure, where the material as PBC in all directions. 
b) Shows how the same Kagome lattice behaves but with PBC only in the horizontal direction. 
c) A simplified representation of how the material behaves at an edge under OBC, exhibiting similarities with the SSH model. 
}}
    \label{fig:B-E-C}
\end{figure}

{\color{black}

In order to analyze our HOTI model, which is a 2D material whose topological states appear in 0D, it is necessary to employ a model with OBC in all directions. One of the simplest solutions is illustrated in Fig.~\ref{fig:B-E-C}, which serves as a guide to the results presented in the main text. We first consider a bulk model, where the system contains periodic boundary conditions (PBC) in all directions; then a model that captures the edge states, where periodicity is kept only along one direction; and finally a fully finite model, with OBC in both directions, which allows the corner states of the material to emerge. For this reason, we introduce the Su--Schrieffer--Heeger (SSH) model. Although this model is not identical to the Kagome lattice, it can be regarded as a simplified representation of one of the edges of the triangular geometry shown in Fig.~\ref{fig:fig0}, due to the similarities shared by both systems.

It is worth noting that another way to study corner states in these types of materials is to consider nanodisks directly~\cite{EzawaPRB2007}, which naturally implement OBC in all directions while preserving the original Kagome structure with three sites per unit cell.  These models are currently under development for future investigations and are therefore not presented in this study.
}

\subsubsection{Bulk Hamiltonian}\label{AppendixA.3a}

\begin{figure}[h!]
    \centering
    \includegraphics[width=0.55\linewidth]{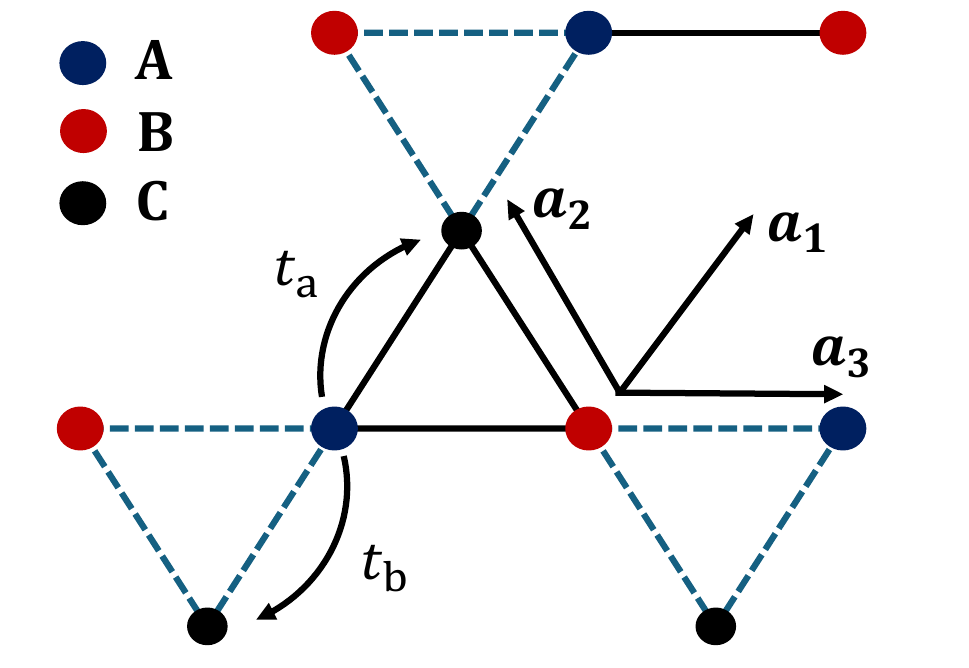}
    \caption{Breathing Kagome lattice geometry: Schematic of a two-dimensional breathing Kagome lattice, a distinct geometry formed by corner-sharing triangles with sites labeled A, B, and C. Each adjacent pair of triangles is inverted relative to the other, producing a periodic pattern of hexagons interlaced with triangles. Gray, black, and blue dots mark the lattice sites. Solid and dashed lines indicate two inequivalent nearest-neighbor hoppings; when their magnitudes differ, inversion symmetry (IS) is broken. Arrows denote the hopping direction, which depends on the specific model under consideration.}
    \label{fig:FigS2}
\end{figure}
\noindent The kagome model describes a two-dimensional system that can host corner states or edge states, corresponding to higher-order topological insulators (HOTIs) and topological insulators (TIs), respectively. Each case is characterized by its own topological index. We employ a tight-binding model with nearest-neighbor hopping \( t_a \) between the atoms in the unit cell (A, B, C), defined by the lattice vectors \( \mathbf{a}_i \). We also include hopping to atoms in adjacent unit cells, which are still nearest neighbors but located in the opposite direction,
\begin{equation}
H(\mathbf{k}) =
\begin{pmatrix}
0 &
t_{a}\, e^{i\mathbf{k} \cdot \mathbf{a}_3} + t_{b}\, e^{-i\mathbf{k} \cdot \mathbf{a}_3} &
t_{a}\, e^{i\mathbf{k} \cdot \mathbf{a}_2} + t_{b}\, e^{-i\mathbf{k} \cdot \mathbf{a}_2} \\
t_{a}\, e^{-i\mathbf{k} \cdot \mathbf{a}_3} + t_{b}\, e^{i\mathbf{k} \cdot \mathbf{a}_3} &
0 &
t_{a}\, e^{i\mathbf{k} \cdot \mathbf{a}_1} + t_{b}\, e^{-i\mathbf{k} \cdot \mathbf{a}_1} \\
t_{a}\, e^{-i\mathbf{k} \cdot \mathbf{a}_2} + t_{b}\, e^{i\mathbf{k} \cdot \mathbf{a}_2} &
t_{a}\, e^{-i\mathbf{k} \cdot \mathbf{a}_1} + t_{b}\, e^{i\mathbf{k} \cdot \mathbf{a}_1} &
0
\end{pmatrix}.
\end{equation}

\noindent The lattice vectors are defined as:
\begin{equation}
\mathbf{a}_1 = \frac{1}{2}\left(1, \sqrt{3}\right) a_0, \quad
\mathbf{a}_2 = \frac{1}{2}\left(-1, \sqrt{3}\right) a_0, \quad
\mathbf{a}_3 = \left(1, 0\right) a_0.
\end{equation}

\noindent Inversion symmetry can be broken by setting different values for \( t_a \) and \( t_b \). When \( t_a = t_b \), the system is in a topological Dirac semimetal phase.  
We first briefly discuss the topology of the system without trimerization (\(t_a = t_b\)), referred to as the SM-TM phase. In this case, the system is inversion symmetric, and the bulk topological index, the \(\mathbb{Z}_2\) topological invariant \(\nu\), can be obtained from the parity eigenvalues \(\xi_{2m}(\Gamma_{1-4})\) of the \(2m\)-th occupied bands, with \(m = 1, 2, 3\), at the four time-reversal-invariant (TRI) points~\cite{FuPRB2007}.  
For the kagome lattice, the parity operator at the TRI points is spin-independent and can be expressed as a \(3 \times 3\) matrix.  
The \(\mathbb{Z}_2\) index for the \(m\)-th Kramers pair is given by 
\begin{equation}
    (-1)^{\nu_m} = \prod_i \xi_{2m}(\Gamma_i).
\end{equation}

\noindent By explicitly calculating the Bloch wave functions, we find \(\nu_m = 1\) for all \(m\), hence the overall \(\mathbb{Z}_2\) index is always \(\nu = 1\)~\cite{BolensPRB2019, GuoPRB2009}.~In the HOTI phase, the breathing kagome lattice exhibits three mirror symmetries: one with respect to the \(x\)-axis and two others related to lines obtained by rotating the \(x\)-axis by \(2\pi/3\).  
The polarization along a given axis \(x_i\) is defined as
\begin{equation}
p_i = \frac{1}{S} \int_{\text{BZ}} A_i \, d^2\mathbf{k},
\end{equation}
where \(A_i = -i \langle \psi | \partial_{\mathbf{k_i}} | \psi \rangle\) denotes the Berry connection for \(x_i = x, y\), and \(S\) corresponds to the Brillouin zone area.  
It is important to note that \(p_x\) is only well defined modulo 1, since it changes by an integer under gauge transformations.  
The polarizations along the other two mirror directions can be expressed as
\begin{equation}
p_{\pm} = -\frac{p_x}{2} \pm \frac{\sqrt{3}}{2} p_y.
\end{equation}
By combining the \(C_3\) rotational symmetry with the mirror operations, one can introduce the following invariant:
\begin{equation}
P_3 = p_x^2 + p_+^2 + p_-^2 = \frac{3}{2} \left( p_x^2 + p_y^2 \right).
\end{equation}
\noindent This construction provides a characterization of the transition between the trivial insulating state and the HOTI phase, which is controlled by the ratio \(t_a / t_b\). 
{\color{black}In the literature, it has been reported that the topological phase of this material lies within the interval 
$-1 < t_a/t_b < 1/2$. However, the bulk topological index is actually non-trivial over the wider range 
$-1 < t_a/t_b < 1$. The apparent reduction of the topological window arises from the overlap between bulk 
states and corner states in finite geometries (e.g., nanodisks), which makes it difficult to clearly 
distinguish the corner modes from the rest of the spectrum~\cite{KempkesNatureMat2019, EzawaPRL2018}. 

Since in this work we focus on the SSH-type version of the model, it is appropriate to consider the full 
interval $[-1, 1]$ as the topological region, as it corresponds to the same range in which the SSH model 
exhibits its topological phase (highlighting the close similarity between both models).
}
\begin{figure}[h!]
    \centering
    \includegraphics[width=0.55\linewidth]{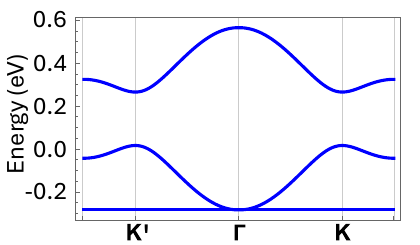}
    \caption{{ \color{black}We show bulk band structure of the Kagome lattice model used for the HHG calculations in Fig.~\ref{fig:fig6}. 
The hopping parameters for the HOTI phase are $t_a = 0.0037~\text{a.u.}$ ($\sim 0.10$ eV) and 
$t_b = 0.0067~\text{a.u.}$ ($\sim 0.183$ eV), for which the ratio $t_a/t_b \sim 0.55$ corresponds 
to the HOTI phase. In the trivial phase, the parameters are inverted, $t_b \rightarrow t_a$ and 
$t_a \rightarrow t_b$, resulting in a ratio $t_a/t_b = 1.83$, characteristic of the trivial regime. 
As in the results shown in Fig.~\ref{fig:fig5}a,c), the band structures are identical in shape but possess 
different topological indices. In addition, the model exhibits a band gap of $\varepsilon_g^{(0)} \sim 0.25~\text{eV}$.
}}
    \label{fig:FigS4}
\end{figure}


{\color{black}
\subsubsection{Edge Hamiltonian}\label{AppendixA.3b}
}
{\color{black}We consider a finite-width Kagome ribbon composed of \(N_x\) unit cells along the periodic \(x\)-direction and \(N_y\) unit cells along the finite \(y\)-direction. Each unit cell contains three inequivalent basis sites (A, B, C). This geometry yields a zigzag nanoribbon with open boundary conditions along the \(y\)-direction. The tight-binding Hamiltonian includes only nearest-neighbor (NN) interactions and, in real space, reads:
}
\begin{equation}
    H_0 =  \sum_{\langle i,j\rangle} t_{ij}\, c_i^\dagger c_j +\text{h.c}
    \label{eq_K:H_0}
\end{equation}

{\color{black}Where $t_{ij} \in \{ t_a, t_b\}$. Because the ribbon is periodic only along \(x\), we perform a partial Bloch transformation
}
\begin{equation*}
 c_{m,n,s} = \frac{1}{\sqrt{N_k}} 
    \sum_{k_x} e^{i k_x x_{mns}} c_{k_x,n,s}
    \label{eq_K:TF_partial}
\end{equation*}

{\color{black}$s$ labels the A, B, C sites within that unit cell. which yields a \(kx-\)dependent Hamiltonian matrix Eq.\ref{eq_H:H_k}
The first (\(h_{i,i+1}\)) and second (\(h_{i,i+2}\)) superdiagonals correspond to nearest-neighbor (NN) hoppings. In this ribbon geometry, the NN terms occupy the first two diagonals because the underlying unit cell contains three atomic sites, unlike the hexagonal (two-site) case where NN hoppings appear only in the first superdiagonal. The explicit NN blocks are
}
\begin{equation}
    h_{i,i+1} \;\rightarrow\; 
    \big( t_a e^{i k_x a/2},\;\;
    t_a e^{-ik_x a} + t_b e^{ik_x a},\;\;
    t_b e^{-ik_x a/2} \big),
    \label{eq_K:1st_diagonal}
\end{equation}
{\color{black} and the second superdiagonal, which also contributes to NN hopping within the enlarged three-site unit cell, is given by
}
\begin{equation}
    h_{i,i+2} \;\rightarrow\; 
    \big( t_a e^{-ik_x a/2},\;\; 
    t_b e^{ik_x a/2},\;\; 
    0 \big).
    \label{eq_K:2nd_diagonal}
\end{equation}
 {\color{black} As in the bulk Hamiltonian, there is no main diagonal (i.e., no onsite potential), so the diagonal blocks remain zero.
Finally, the lower-diagonal blocks are obtained from the Hermitian conjugates of the corresponding upper-diagonal terms:
 }
\begin{equation}   
    h_{i,i-1} = h_{i-1,i}^\dagger, \qquad
    h_{i,i-2} = h_{i-2,i}^\dagger.
    \label{eq_K:h.c.}
\end{equation}
\noindent{\color{black}The full Hamiltonian takes the block form:
}
\begin{equation}
    H(k_x)=
    \begin{pmatrix}
    0 & t_a e^{i k_x a/2} & t_a e^{-ik_x a/2} & 0 & \cdots \\
    h_{1,2}^\dagger & 0 & t_a e^{-ik_x a} + t_b e^{ik_x a} &  t_b e^{ik_x a/2} & \cdots \\
    h_{1,3}^\dagger & h_{2,3}^\dagger & 0 & t_b e^{-ik_x a/2} & \cdots \\
    0 & h_{2,4}^\dagger & h_{3,4}^\dagger & 0 & \cdots \\
    \vdots & \vdots & \vdots & \vdots & \ddots
    \end{pmatrix}
\end{equation}
{\color{black}This formalism directly generalizes to a Kagome ribbon of arbitrary size $N_x \times N_y$, with periodic boundary conditions along $x$ and open boundary conditions along $y$. Since each unit cell contains three sites, the full Hamiltonian has dimension $3 N_x N_y$.
}

{\color{black}
\subsubsection{SSH Hamiltonian}\label{AppendixA.3c}
}

\begin{figure}[h!]
    \centering 
    \includegraphics[width=0.95\textwidth]{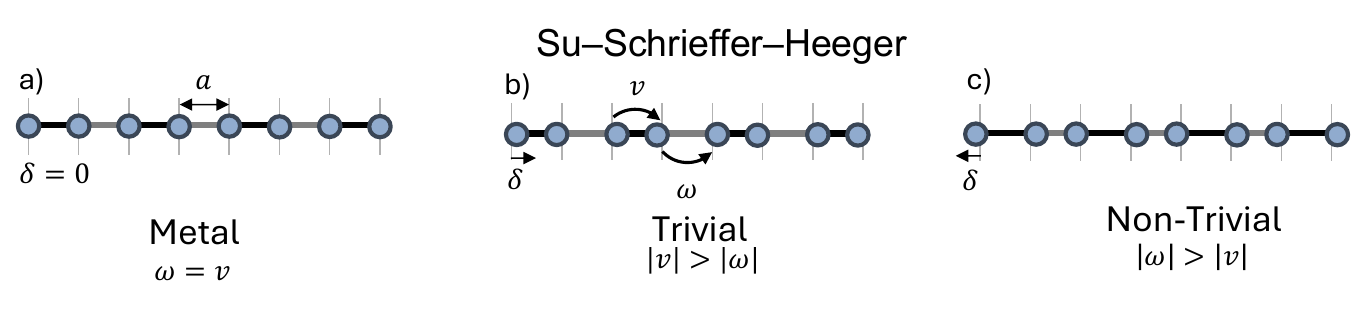}
    \caption{{\color{black}a--c) Represent the SSH chain in its metallic, trivial, and topological phases, respectively. Specifically, a) shows the definition of \(a\) in the metallic phase, where a uniform lattice is observed. 
In b--c), the lattice becomes distorted due to a displacement \(\delta\), causing the atoms to organize into pairs (dimerization). However, these two configurations correspond to different topologies under OBC. 
b) shows how the lattice dimerizes such that all atoms pair up, corresponding to a trivial topology. 
c) shows the chain dimerizing in the opposite direction, causing the first and last atoms of the lattice to remain unpaired. This behavior generates a non-trivial topology characterized by degenerate zero-energy states.
}}
    \label{fig:SSH}
\end{figure}

{\color{black}The Su--Schrieffer--Heeger (SSH) model consists of a chain with two sites per unit cell. It can be viewed under PBC, where the material still preserves translational symmetry; however, it can also be studied as a finite-length chain, in which case we have OBC in all directions. By considering an even number of sites $N$, the SSH model consists of $n = N/2$ primitive cells, each containing two lattice sites labeled $s = A, B$. The nearest-neighbor hoppings are described by $v$ and $w$.}
 \begin{equation}
    H = v \sum_{i}^{N} c^{\dagger}_{A,i}c_{B,i} + w \sum_{i}^{N} c^{\dagger}_{B,i}c_{A,i+1} + \text{h.c.}
 \end{equation}
 {\color{black}The hoppings $v$ and $w$ are defined in terms of two parameters: $a$, which represents the distance between atoms in the metallic case $\delta = 0$, and $\delta$, which represents the alternating displacement of atoms in the lattice (dimerization). This has effects analogous to trimerization in the case of the Kagome lattice, controlling the band gap of the material.}
\begin{align}
    v =& - \exp[-(a-2\delta)] \\
    w =& - \exp[-(a+2\delta)]
\end{align}

{\color{black}Since the system considers only NN, we will have only one diagonal above the main one
}

\begin{equation}
    h_{i,i+1} \;\rightarrow\; 
    \big( v,\;\;
    w \big)
    \label{eq_S:1st_diagonal}
\end{equation}

{\color{black}Finally, the lower diagonal blocks are given by the Hermitian conjugates of the corresponding
upper blocks,}

\begin{equation}   
    h_{i,i-1} = h_{i-1,i}^\dagger,
    \label{eq_S:h.c.}
\end{equation}

{\color{black}The Hamiltonian for a finite chain is}
\begin{equation}
    H =
\begin{pmatrix}
0 & v & 0 & 0 & \cdots  \\
v & 0 & w & 0 & \cdots  \\
0 & w & 0 & v & \cdots  \\
0 & 0 & v & 0 & \cdots  \\
\vdots & \vdots & \vdots & \vdots & \ddots\\

\end{pmatrix}.
\label{eq_S:H}
\end{equation}


\section{HHG calculation}\label{AppendixB}
\subsection{Time Dependent Density Matrix on a Tight Binding Basis}\label{AppendixB.1}
It is possible to calculate the high-harmonic generation (HHG) spectra using the time dependent density matrix (TDDM), which have been employed in previous works~\cite{SilvaNatPho2019,BaykushevaPRA2021}. In these equations, the system is subjected to a laser field \({\bf E}(t)\), and the dynamics are described by the time evolution of density matrix \({\hat \rho}^{(H)} ({\bf K},t)\) by:
\begin{equation}
i \frac{\partial}{\partial t} \rho_{mn}^{(H)}(\mathbf{K}, t) = \left[ H_0^{(H)}(\mathbf{K} + \mathbf{A}(t)), \rho^{(H)}(\mathbf{K}, t) \right]_{mn} + \mathbf{E}(t) \cdot \left[ \mathbf{D}^{(H)}(\mathbf{K} + \mathbf{A}(t)), \rho^{(H)}(\mathbf{K}, t) \right]_{mn} - i (1 - \delta_{mn}) \frac{\rho_{mn}^{(H)}}{T_2}.
\end{equation}

\noindent The TDDM includes material-specific terms, such as the unperturbed Hamiltonian \(H_0(\mathbf{k})\), which is generally obtained from analytic tight-binding models, (TBM)~(see. Appendix~\ref{AppendixA}). It also incorporates the dipole matrix \( \mathbf{D}_{mn}^{(H)}(\mathbf{k}) = i\langle u_m | \partial_{\mathbf{k}} | u_n \rangle\), where \(|u_m^{(H)}(\mathbf{k})\rangle = e^{-i \mathbf{k} \cdot \mathbf{r}} |\psi_m^{(H)}(\mathbf{k})\rangle\). This corresponds to the Berry connection when \(n = m\) and to the transition dipole matrix elements when \(n \neq m\). Moreover, laser-specific terms are included, such as the electric field \({\bf E}(t)\) and the vector potential \(\mathbf{A}(t)\). Finally, the last term describes the decay of the system back to its ground state, indicating the timescale required to return to the initial condition after laser excitation~\cite{KimJourPhy2023}.
To avoid issues with wavefunction gauge calibration, the Wannier gauge is used in the TDDM~\cite{KimJourPhy2023,KimMDPI2022}. In this gauge, both the Hamiltonian \(H_0^{(W)}\) and the dipole matrix \(\mathbf{D}^{(W)}\) vary smoothly, preventing numerical errors associated with discontinuities in the wavefunction gauge.

\noindent For this purpose, we consider a set of \(M\) Wannier orbitals localized in each cell, defined as
\[
w_m(\mathbf{r} - \mathbf{R}) = \langle \mathbf{r} | \mathbf{R} m \rangle.
\]

\noindent We assume these orbitals form an orthonormal basis, i.e.,
\[
\langle \mathbf{R} m | \mathbf{R}' n \rangle = \delta_{m,n} \, \delta_{\mathbf{R}, \mathbf{R}'}.
\]
\noindent Furthermore, considering a system with \(N_c\) unit cells under periodic boundary conditions, the Bloch functions can be defined as
\begin{align}
|\psi^{(W)}\rangle = \frac{1}{\sqrt{N_c}} \sum_{\mathbf{R}} e^{i \mathbf{k} \cdot (\mathbf{R} + \Delta_n)} |\mathbf{R}  n    \rangle,
\end{align}

\noindent where \(\Delta_n\) is the Wannier center \(\langle \mathbf{0} n | \hat{r} | \mathbf{0} n \rangle\), representing the atomic position within the unit cell. With this, the Hamiltonian in the Wannier gauge is defined as
\begin{align}
    H_{nm}^{(W)}(\mathbf{k}) = \langle \psi_{nk}^{(W)}(\mathbf{k}) | \hat{H} | \psi_{mk}^{(W)}(\mathbf{k}) \rangle = \sum_{\mathbf{R}} e^{i \mathbf{k} \cdot (\mathbf{R} - \Delta_n + \Delta_m)} \langle \mathbf{0} n | \hat{H}_0 | \mathbf{R} m \rangle.
\end{align}

\noindent We note that the basis of wavefunctions used here differs slightly from that employed previously (\(|\psi^{(W)}\rangle = \frac{1}{\sqrt{N_c}} \sum_{\mathbf{R}} e^{i \mathbf{k} \cdot \mathbf{R}} |\mathbf{R} m \rangle\))~\cite{BaykushevaPRA2021,SilvaNatPho2019,VampaPRL2014}. This allows us to manipulate the dipole matrix in this new basis, therefore, by considering highly localized Wannier functions, we can approximate \(\langle \mathbf{0} m | \hat{r} | \mathbf{R} n \rangle \approx \delta_{nm} \Delta_n\). This approximation eliminates transition matrix elements and Berry curvature terms, \(\langle u_m^{(W)}| \partial_{\mathbf{k}} | u_n^{(W)} \rangle = 0\). Moreover, this transformation only affects the transition matrix elements or phase factors~\cite{KimMDPI2022}. Hence, our TDDM becomes
\begin{equation}
    i \frac{\partial}{\partial t} \rho^{(W)}(\mathbf{K}, t) = \left[ H_0^{(W)}(\mathbf{K} + \mathbf{A}(t)), \rho^{(W)}(\mathbf{K}, t) \right].
\end{equation}

\noindent Within this framework, we treat the dephasing time in the eigenstate basis, and the remaining terms are computed in the new basis. After obtaining the time-dependent density matrix, the system's current \(\bf{J}(t)\) can be calculated as
\begin{equation}
    \mathbf{J}(t) = \sum_{mn} \int_{BZ} dK \, \mathbf{P}_{mn}^{(W)}(\mathbf{K} + \mathbf{A}(t)) \, \rho_{nm}^{(W)}(\mathbf{K}, t),
\end{equation}

where 
\begin{equation}
    {\bf P}_{mn}^{(W)}({\bf k}) = \langle \psi_m^{(W)} | \partial_{\bf k}\hat{H}_0^{(W)}({\bf k}) | \psi_n^{(W)} \rangle\
\end{equation}

\noindent It is the momentum matrix element in the Wannier gauge. Finally, the spectral yield of the HHG can be calculated via the Larmor formula, which is proportional to the spectral intensity of the current:
\begin{equation}
    I(\omega) \propto \omega^2 \| \mathbf{J}(\omega) \|^2,
\end{equation}
where a Fourier transform is applied to the time-dependent current. In some cases, for very long times, the integral is limited to finite times by multiplying the current by a windowing function that smoothly reduces it:
\begin{equation}
    \mathbf{J}(\omega) = \frac{1}{\sqrt{2 \pi}} \int_{-\infty}^{\infty} \mathbf{J}(t) e^{i \omega t} dt.
\end{equation}

{\color{black}In our calculation, the electric field is given by  \(E(t) =\partial_t \mathbb{A}(t)\), where the vector potential is defined as
}
\begin{equation}
    \mathbf{A}(t)
= \frac{E_{0}}{\omega_{0}}\, f(t)
\left(- \frac{1}{\sqrt{1+\epsilon^{2}}}
\sin\!\big(\omega_{0}(t - t_{0}) - \phi_{0}\big)\, \hat{\mathbf{e}}_{x}
+ {\frac{\epsilon}{\sqrt{1+\epsilon^{2}}}}\,
\cos\!\big(\omega_{0}(t - t_{0}) - \phi_{0}\big)\, \hat{\mathbf{e}}_{y}\right).
\end{equation}

{\color{black}where \(E_0\) is the electric field peak strength, \(\omega_0\) is the laser angular frecuency, \(\phi_0\) is the carrier-envelope phase, \(\epsilon\) is the ellipticity of the laser, and \(f(t)\)  is the laser pulse envelope. We use Gaussian envelope \(f(t) = \exp(-4 \log 2 \frac{\left(t-t_0\right)^2}{t^2_{\sigma}})\)
where \(t_0\) is the center of the pulse and \(t_{\sigma}\) is the full width at half maximum (FWHM) of the laser.
}

{\color{black}
\subsection{Time-dependent Schrödinger Equation (TDSE)}\label{AppendixB.2}

For the finite chains, high-harmonic spectra may be obtained from the dipole, the acceleration, or the current~\cite{pooyan2025harmonic, jurss2022topological, drueke2021high}. In the length gauge, the diagonal elements of the Hamiltonian \(H_{jj}\) need to be replaced by, 
}
\begin{equation}
    H_{jj} \to E(t)x_j
\end{equation}

{\color{black}thus the Hamiltonian matrix reads}
\begin{equation}
    H(t) =
\begin{pmatrix}
E(t)x_1 & v & 0 & 0 & \cdots  \\
v & E(t)x_2 & w & 0 & \cdots  \\
0 & w & E(t)x_3 & v & \cdots  \\
0 & 0 & v & \ddots & \cdots  \\
\vdots & \vdots & \vdots & \vdots & E(t)x_n\\

\end{pmatrix}.
\label{eq_S:H(t)}
\end{equation}

{\color{black}In this paper, to comuted the harmonic spectrum, we first evaluate the electron position expectation value
}
\begin{equation}
    \mathbf{X}(t) = \sum_{i=0}^{N/2-1} \sum_{j=1}^N \langle \Psi_i^j(t) | x_j | \Psi_i^j(t) \rangle,
\end{equation}
{\color{black}
where $x_j$ is atomic-site positions and the $\Psi(t)$ is obtaind from wave functions propagation in time using the CrankNicolson approximant to the time-evolution operator
}
\begin{equation}
    \exp(-iH(t)\Delta t) = \frac{1-iH(t) \Delta t/2}{1+iH(t) \Delta t/2}+\mathcal{O}(\Delta t^3).
\end{equation}
{\color{black}
Finally, in the semiclassical treatment, and assuming the rmitters are uncorrelated, the spectrum of the radiated light $I(\omega)$ is proportional to the square magnitude of the Fourier-transformed dipole acceleration
}
\begin{equation}
    I(\omega) = |{\rm FFT}[\ddot{\mathbf{X}} (t)]|^2
\end{equation}

{\color{black}
\subsection{Additional Results for the LOTI}\label{}
}
{\color{black}
In this section, we present some results that were omitted from the main text but are nonetheless relevant.
In Fig.~\ref{fig:Haldane_and_Kane--MeleBulk_vs_Edge}, the same results as in Fig.~\ref{fig:fig2} are shown, but now with a different configuration, allowing for a more convenient comparison between the trivial and topological phases in their respective channels. In the second case, Fig.~\ref{fig:Haldane_Helicty} shows the LP(+) and LP(--) decompositions, which rotate to the right and left, respectively, generated from a linearly polarized laser. These spectra are used to obtain the values reported in Table~\ref{tab:cd_helicity}.
}

\begin{figure} [htbp]
    \centering
    \includegraphics[width=0.8\linewidth]{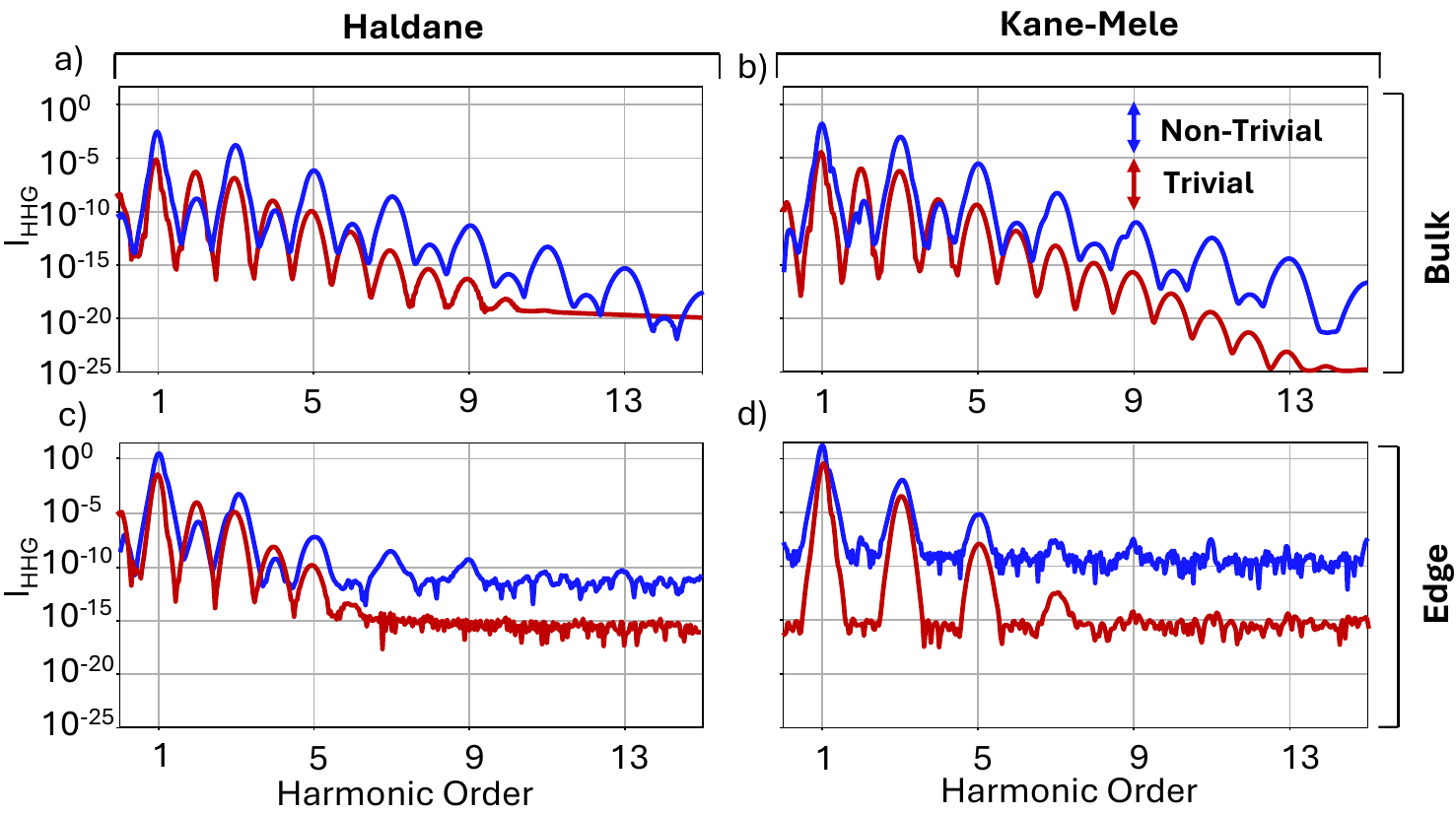}
    \caption{{\color{black} Comparison of the high harmonic generation (HHG) spectra produced from {\it bulk} and {\it edge} states in LOTIs:
Panels~a) and~b) correspond to the {\it bulk emission} for the 2D Haldane and Kane--Mele models, respectively, while panels~c) and~d) show the corresponding {\it edge emission}.
Within each panel, we compare the HHG spectra for the {\it trivial} (red line) and {\it non-trivial} (blue line) topological phases.
All simulations (bulk and edge) use the same laser and topological material parameters as in Fig.~\ref{fig1}.
Edge states are computed for a one-dimensional zig-zag hexagonal strip (see Appendix~\ref{AppendixA.1b}).
Both the Haldane and Kane--Mele models include next-nearest-neighbor terms, whereas only the Kane--Mele model incorporates spin (up and down).
Overall, the HHG signal from {\it edge states} is significantly enhanced compared to the {\it bulk states}, particularly in the non-trivial topological phase.}}
    \label{fig:Haldane_and_Kane--MeleBulk_vs_Edge}
\end{figure}

\begin{figure}
    \centering
    \includegraphics[width=0.75\linewidth]{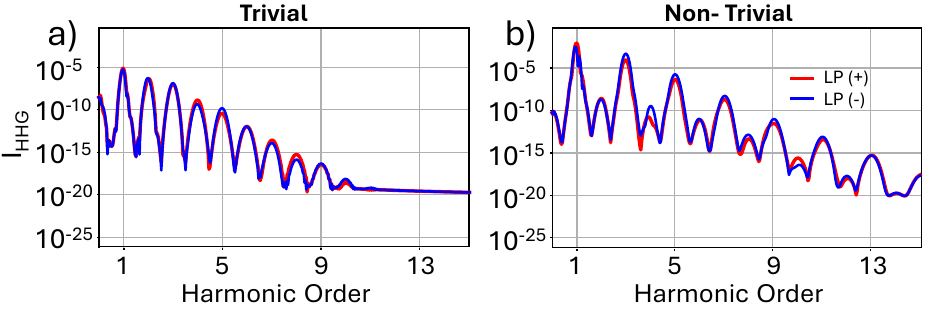}
\caption{{\color{black} Comparison of the high harmonic generation (HHG) spectra driven by a linearly polarized laser for the 2D Haldane model.
Panels~a) and~b) correspond to the {\it trivial} and {\it non-trivial} topological phases, respectively.
Within each panel, we compare the circular decomposition of the emitted harmonics into right-handed ({\it +}) and left-handed ({\it -}) rotating components generated from the linearly polarized driving field.} }
   \label{fig:Haldane_Helicty}
\end{figure}

\newpage

{\color{black}
\subsection{HHG result for phase transition in SSH}\label{AppendixB.3}
}
\begin{figure}[h!]
    \centering 
    \includegraphics[width=1\textwidth]{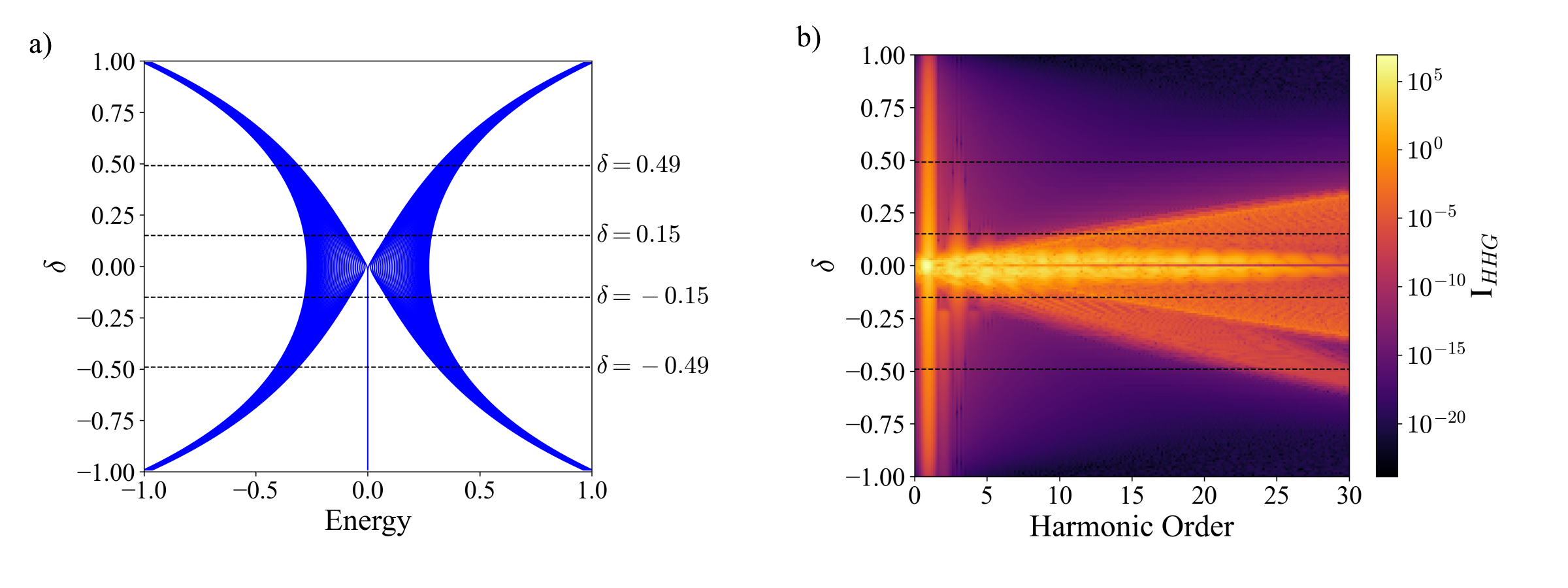}
    \caption{{\color{black}Energy spectrum and high-harmonic spectra for the chain under OBC. 
a) Energy spectrum as a function of the parameter $\delta$, obtained from the diagonalization of Eq.~\ref{eq_S:H}. 
Positive and negative values of $\delta$ exhibit similar behavior, except that negative $\delta$ always hosts states at the Fermi level, corresponding to corner modes of the chain. 
b) High-harmonic emission for the same $\delta$-scan. 
The dashes lines in both panels mark the values $\delta=\pm 0.15$ and $\delta=\pm 0.49$, which are discussed in the main text. 
Laser parameters are identical to those used in Fig.~\ref{fig1}.
}}
\label{fig:figScan}
\end{figure}

{\color{black}
In Fig.~\ref{fig:figScan}a) the energy spectrum of the SSH model is shown as a function of the parameter $\delta$, which controls the dimerization of the system and produces a progressive opening of the chain's band gap. The plot also distinguishes between the trivial ($\delta>0$) and topological ($\delta<0$) regions; the latter is characterized by topological states or zero-energy modes (visible only under OBC). In Fig.~\ref{fig:figScan}b) the HHG results are shown for the same systems depicted in Fig.~\ref{fig:figScan}a). Since the only difference between the topological and trivial regions is the presence of edge states at the ends of the chain, it is expected that both regions exhibit generally similar spectra.

Very close to $\delta = 0$, where the topological transition occurs, an intense horizontal band appears, indicating strong emission near the metallic phase or in regimes with a very small band gap (not necessarily related to topology, as this behavior is symmetric). However, starting from the 5th harmonic, a progressive ``fan-shaped'' enhancement emerges as a function of $\delta$. This enhancement is completely asymmetric, revealing that the topological region exhibits stronger emissions for high-order harmonics.

This explains why the two topological phases in Fig.~\ref{fig:fig6}c) and Fig.~\ref{fig:fig6}f) behave differently. For values around $\delta \approx 0.15$, differences between the phases become visible starting from the 5th harmonic. However, for values near $\delta \approx 0.49$, the difference between the phases appears only above the 20th harmonic—a region not explored in the main text.

From these results we conclude that it is indeed possible to extract signatures of topology through HHG, provided that models with open boundary conditions are employed, as these are essential for the existence of topological states. However, such signatures are not universal, because the generation of harmonics depends not only on topology but also on various features of the band structure, such as the band-gap size, spin, and other electronic details. This implies that these signatures will not be fixed even within the same system. In the model analyzed here, for instance, the enhancement of high-harmonic emission depends not only on being in a topological phase but also on the value of the band gap, which causes the asymmetry of the ``fan-shaped'' pattern to shift toward even higher-order harmonics as the band gap increases.
}

{\color{black}
\subsection{HHG Results for Bi$_2$Se$_3$ and Graphene}\label{AppendixB.5}
}

{\color{black}
In this section, we present additional HHG results for the 3D topological insulator Bi$_2$Se$_3$ and the trivial material graphene. In particular, in Fig.~\ref{fig:ReplyFig0} we compare the harmonic emission generated from the bulk and topological surface states of Bi$_2$Se$_3$, as well as the response of the topological surface states of Bi$_2$Se$_3$ against graphene in Fig.~\ref{fig:ReplyFig1}. These comparisons further illustrate the enhancement of HHG associated with topological states, and reinforce the conclusion that the intensity yield of the harmonic emission serves as a reliable indicator for distinguishing topologically nontrivial states from trivial ones, consistent with the results reported in the main text for both LOTI and HOTI materials.
}

\begin{figure}[htbp]
\begin{center}
\includegraphics[width=0.45\linewidth]{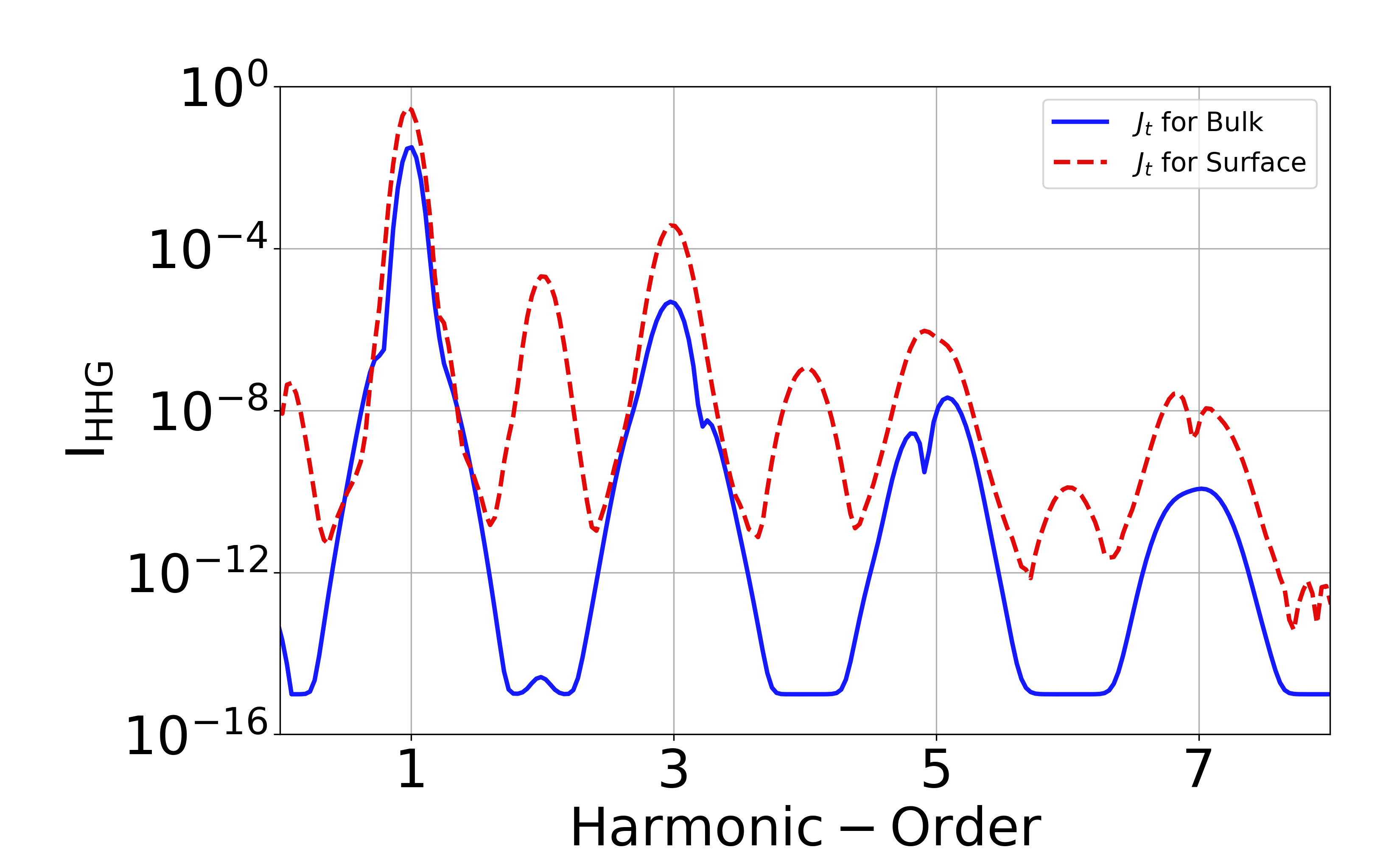}
    \caption{High-harmonic generation produced by linearly polarized light from the 3D topological insulator Bi$_2$Se$_3$. The blue solid line and red dashed line depict the HHG spectra emitted from the 3D topological bulk and 2D topological surface states, respectively. The laser parameters are field strength $E_0 = 0.0007~\mathrm{a.u.}$ ($I_0 = 1.7 \times 10^{10}~\mathrm{W/cm}^2$), photon energy $\hbar\omega_0 = 0.013~\mathrm{a.u.} \approx 0.35~\mathrm{eV}$ ($\lambda = 3.54~\mu\mathrm{m}$), and a Gaussian envelope with a pulse duration of 7 cycles at full width at half maximum (FWHM). The bulk and surface model parameters are the same as those reported in Ref.~\cite{BaykushevaPRA2021, BaykushevaNanoLetters2021, MaoPRB2011}. We stress that the harmonic emission from the surface states exhibits an enhancement of between two and three orders of magnitude compared to the bulk contribution.}
    \label{fig:ReplyFig0}
\end{center}
\end{figure}

\begin{figure} [htbp]
\begin{center}
    \includegraphics[width=0.45\linewidth]{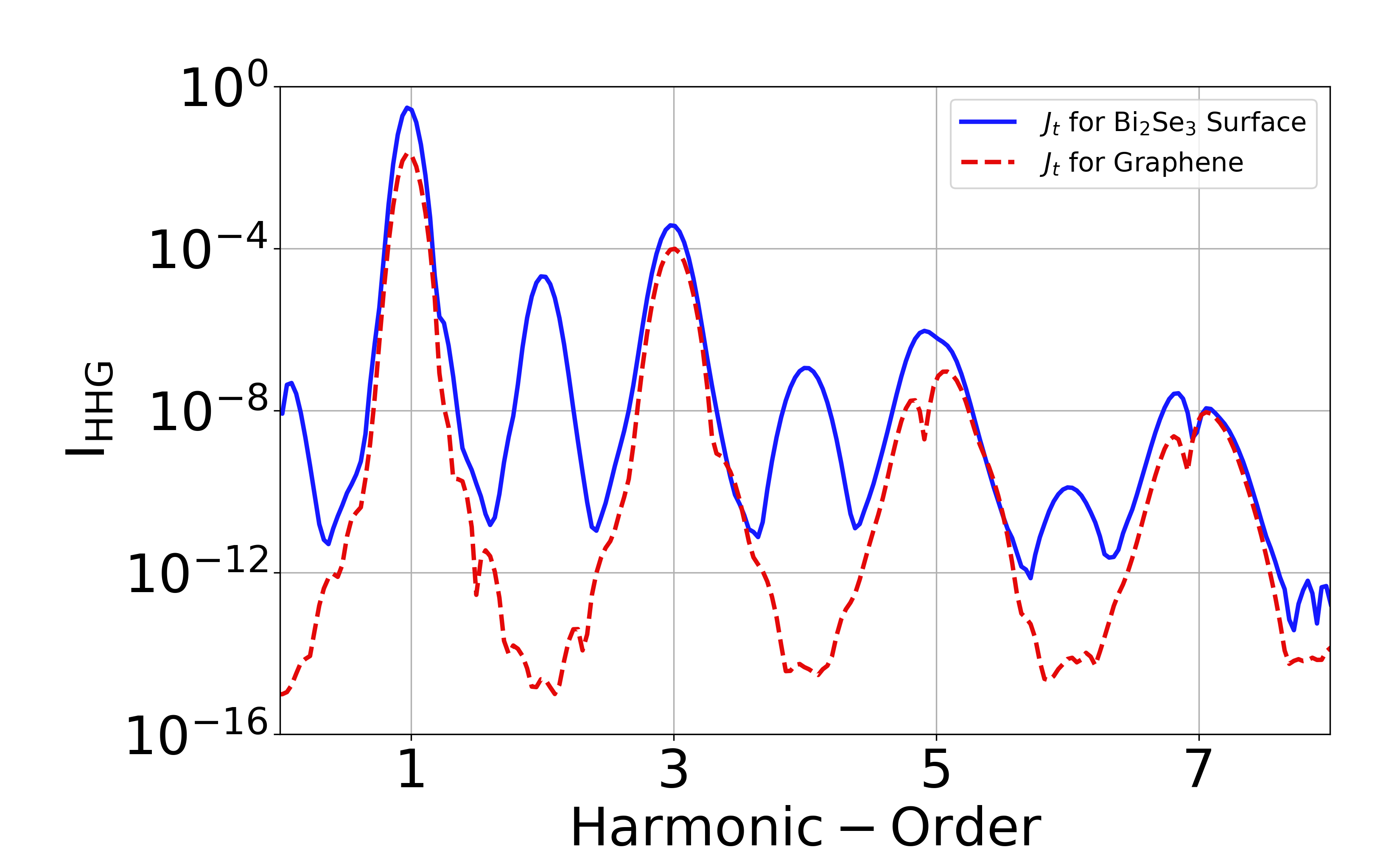}
    \caption{High-harmonic generation produced by linearly polarized light from a topological insulator and a trivial material. The blue solid line and red dashed line represent the HHG spectra emitted from the 2D topological surface states of Bi$_2$Se$_3$ and the 2D trivial ``bulk'' states of graphene, respectively. For graphene, we used the Haldane-like model with nearest-neighbor hopping $t_1 = 0.1029~\mathrm{a.u.}$ ($\approx 2.80~\mathrm{eV}$), next-nearest-neighbor hopping $t_2 = 0~\mathrm{a.u.}$, on-site potential ratio $M_0/t_2 = 0$, and magnetic flux $\phi_0 = 0$. For Bi$_2$Se$_3$, we focus exclusively on the topological surface states, using the same model parameters reported in Ref.~\cite{BaykushevaPRA2021, BaykushevaNanoLetters2021, MaoPRB2011}, with intralayer hopping parameters $B_0 = 0.0164~\mathrm{eV}$ and $B_{11} = 0.1203~\mathrm{eV}$. The laser parameters are the same as those used in Fig.~\ref{fig:ReplyFig0}. A clear enhancement is observed for the harmonic emission from the topological surface states of Bi$_2$Se$_3$ compared to the trivial graphene response.}
    \label{fig:ReplyFig1}
\end{center}
\end{figure}

\newpage

\bibliographystyle{apsrev4-2}
\bibliography{apssamp.bib}

\end{document}